\documentclass{emulateapj}

\def\brfrac#1#2{\left(\dfrac{#1}{#2}\right)}
\def\dfrac#1#2{{\displaystyle\frac{\mathstrut #1}{#2}}}

\slugcomment{Received 2011 April 26; accepted 2011 July 3}

\shorttitle{NIR 
Reverberation by AGN Dusty Clumpy Tori}
\shortauthors{Kawaguchi and Mori}

\begin{document}

\title{Near-Infrared 
Reverberation by Dusty Clumpy 
Tori in Active Galactic Nuclei}

\author{Toshihiro Kawaguchi and Masao Mori}
\affil{Center for Computational Sciences, University of Tsukuba, 
Tsukuba, 
Ibaraki 305-8577, Japan}

\email{kawaguti@ccs.tsukuba.ac.jp}

\begin{abstract}
According to recent models,
the accretion disk and black hole in active galactic nuclei 
are surrounded by a clumpy torus.
We investigate the NIR flux variation of the torus in response 
to a UV flash for 
various geometries. 
Anisotropic illumination by the disk and 
the torus self-occultation contrast our study with earlier works.
Both the waning effect of each clump and the torus self-occultation 
selectively reduce the emission from  the region with a short delay.
Therefore, the 
NIR delay depends on the 
viewing angle (where a more inclined angle leads to 
a longer delay) and the 
time response shows 
an asymmetric profile with a negative skewness, 
opposing to the results for optically thin tori. 
The range of the computed delay coincides with 
the observed one, suggesting 
 that the viewing angle is 
primarily responsible for the scatter of the observed delay. 
We also propose 
that the red NIR-to-optical color 
of type-1.8/1.9 objects is caused by not only 
the dust extinction but also the intrinsically red color. 
Compared with the modest torus thickness, 
both a thick and a thin tori display the weaker NIR emission.
A selection bias is thus expected such that NIR-selected 
AGNs tend to possess moderately thick tori.
A thicker torus shows 
a narrower and more heavily 
skewed time profile, 
while a thin torus produces a rapid response.
A super-Eddington accretion rate leads to a much weaker 
NIR emission due to the disk self-occultation and 
the disk truncation by the self-gravity.
A long delay is expected from an optically thin 
and/or a largely misaligned torus.
A very weak NIR emission, such as in hot-dust-poor active nuclei, 
can arise from a geometrically thin torus, a super-Eddington 
accretion rate or a slightly misaligned torus.
\end{abstract}


\section{Introduction}

Active galactic nuclei (AGNs) are powered by gas accretion onto
supermassive black holes (BHs) at the center of each galaxy.
A variety of observations suggest that
the accretion disk and the BH are surrounded by an optically and
geometrically thick torus
 (Telesco et al. 1984; 
 Antonucci \& Miller 1985; 
 Miller \& Goodrich 1990;  
 Radovich et al. 1999).
Since the 
torus potentially plays a role of a gas reservoir 
for the accretion disk, 
its nature, 
such as the structure, 
the size and the mass, 
has long been investigated 
(Pier \& Krolik 1992, 1993; 
 Fukue \& Sanbuichi 1993; 
Granato \& Danese 1994; 
Efstathiou \& Rowan-Robinson 1995
Beckert \& Duschl 2004; 
 Mor et al. 2009). 

A large geometrical thickness of the torus 
revealed by various observations
(Antonucci 1993; 
 Pogge 1989; 
 Wilson \& Tsvetanov 1994; Schmitt \& Kinney 1996) 
 indicates
that numerous dusty clumps, 
rather than a smooth mixture of gas and dust, 
constitute the torus with a large clump-to-clump velocity 
dispersion $\sim \! 100\,\mathrm{km\, s}^{-1}$ 
(Krolik \& Begelman 1988; Wada \& Norman 2002; H\"{o}nig \& Beckert 2007).
Temperature of clumps is less than a critical 
temperature $T_{\rm sub} \sim 1500$\,K above which 
dust grains 
are sublimated (Barvainis 1987).
Infrared (IR) emission and absorption features provide unique opportunities 
to probe the clumpy torus (Nenkova et al. 2002, 2008; 
Dullemond \& van Bemmel 2005; H\"{o}nig et al. 2006; 
Geballe et al. 2006; Shirahata et al. 2007; 
Ibar \& Lira 2007; 
Schartmann et al. 2008; 
Deo et al. 2011).

Clumps are heated by illumination 
from the central accretion disk, 
and closer clumps 
have higher temperature. 
Inner edge of the dusty torus is determined by the sublimation 
process so that clumps' temperature equals $T_{\rm sub}$ there, 
and radiates at 
Near-IR (NIR) 
as "3$\mu$m bump" 
(Rees et al. 1969; Neugebauer et al. 1979; 
Edelson \& Malkan 1986; 
Kobayashi et al. 1993).
Based on the energy balance of the clump closest to the BH, 
Barvainis (1987) derived the innermost radius of the torus
(dust sublimation radius, denoted as $R_{\rm sub,0}$ in this
study):  
\begin{equation}
R_{\rm sub,0} = 0.13 \brfrac{L_{\rm UV}}{10^{44} \, \mathrm{erg}\,\mathrm{s}^{-1}}^{0.5}
 \brfrac{T_{\rm sub}}{1500 \,\mathrm{K}}^{-2.8}
 \brfrac{a}{0.05 \,\mu \mathrm{m}}^{-0.5} {\rm pc},
 \label{eq:bar87}
\end{equation}
where $L_{\rm UV}$ 
and $a$ are UV luminosity  
and the size of dust grains, respectively.

Indeed, NIR emission 
from type-1 AGNs lags behind optical variation by an order 
of a month
(Clavel et al. 1989; Glass 1992, 2004; 
Nelson 1996; Oknyanskij et al. 1999; 
Minezaki et al. 2004; Suganuma et al. 2004).
Moreover, the luminosity dependency of the time lag also 
coincides with the theoretical prediction as $\propto L_{\rm UV}^{0.5}$
(Suganuma et al. 2006; 
Gaskell et al. 2007). 
However, the NIR-to-optical time lag 
is systematically smaller than the lag predicted from 
Equation (\ref{eq:bar87}) by a factor of $\sim \! 1/3$ 
(Oknyanskij \& Horne 2001; Kishimoto et al. 2007; Nenkova et al. 2008).
To tackle with this conflict, Kawaguchi \& Mori (2010, hereafter Paper I) 
pointed out that the illumination by 
an optically thick disk is inevitably anisotropic, 
which is a fact missing in deriving Equation (\ref{eq:bar87}).
There is a systematic difference between the inclination angle 
at which we observe the disk in type-1 AGNs and the angle 
at which an aligned torus observes the disk.
The effects of the anisotropic 
illumination 
naturally resolve
the puzzle of the systematic deviation of a factor of $\sim \! 1/3$
(Paper I).

In Paper I, we assumed the configuration appropriate for a 
typical type-1 AGN. 
In this study, we investigate the expected characteristics 
of the NIR emission for 
various geometries of the disk, the torus and the observer. 
Anisotropic illumination by the disk and the effect of 
the torus self-occultation contrast our study with earlier works.
The next section 
describes the calculational methods of our model.
Then, properties of NIR emission from an aligned (Section \ref{sec:aligned}) 
and a misaligned (Section \ref{sec:misaligned}) tori are presented.
Finally, we make a summary 
of this study in Section \ref{sec:summary}.

\section{Model Description} \label{sec:model}

We calculate NIR reverberation/echo 
from the inner part of the torus 
in response to a flash of disk illumination. 
The calculational method is essentially the same as Paper I, 
except for the incorporation of the torus self-occultation in this study.
By considering the anisotropy of the disk illumination, we solved 
the inner structure of the torus, and 
explained why the observed time delay of NIR emission 
is systematically shorter than Equation (\ref{eq:bar87}).
A large grain size and/or extinction 
between the torus and the disk are possibly alternative concepts to reduce 
the inner radius of the torus 
(Maiolino et al. 2001; 
Gaskell \& Benker 2007; 
Gaskell et al. 2007; Kishimoto et al. 2007).

While the geometry of the torus, the disk and the observer appropriate for a 
typical type-1 AGN was assumed in Paper I, 
we here investigate how the NIR response differs 
with various possible geometries. 
A variety of type-1 objects with a common $R_{\rm sub,0}$ are compared.
In other words, we compute for objects with the same 
isotropic-equivalent luminosity 
(``luminosity presuming isotropic emission")
in optical/UV.

Barvainis (1992) examined the NIR response due to  
the dust reverberation of an AGN torus. 
Since the inner radius of the torus is an input parameter there, 
the mean time delay of NIR emission 
behind the optical/UV flux variations 
is simply coupled with the assumed inner radius.
On the other hand,
in Paper I and this study, 
both the inner radius and the time delay are output of calculations.
Moreover, he supposed that the whole torus, an ensemble 
of cube-shaped, optically thick clumps, is optically thin.
Since the inner part of an AGN clumpy torus is 
likely optically thick (Appendix), 
we take into account the torus self-occultation as well as the 
anisotropic emission from spherical, optically thick clumps. 
Similarities and differences in the results for the time response 
between our 
and his calculations are discussed in Sections \ref{sec:phi_contri} 
and \ref{sec:qobs}.

In radiative transfer calculations of dusty clumpy tori 
(e.g., H\"{o}nig et al. 2006; Nenkova et al. 2008), 
the isotropic illumination, mainly in optical and UV, is assumed for simplicity.
We take into account the fact that the disk emission is inevitably 
anisotropic.
Then, the broadband color between the optical/UV radiation from the disk 
and the NIR emission from the torus can be computed appropriately.
In Sections \ref{sec:qobs} and  \ref{sec:qmin}, 
we compare our results with theirs.

Below, our calculational method is summarised.

\subsection{Anisotropic Illumination of Disk}

\begin{figure}
\epsscale{0.9}
\plotone{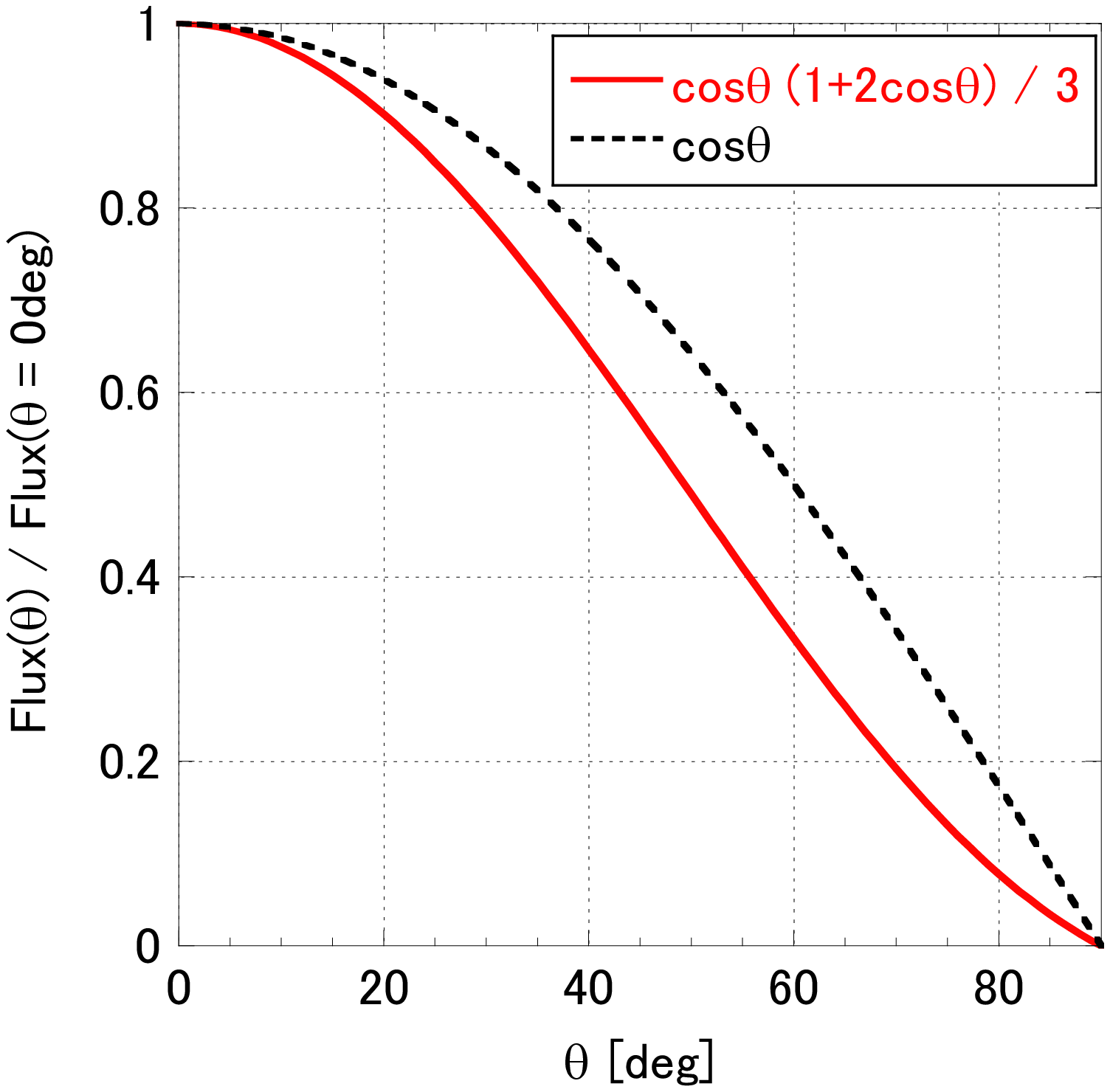}
\caption{
Radiation flux from the disk $F(\theta)$ as a function of 
the polar angle $\theta$, normalised to 
its pole-on value.
Red solid line represents the net $\theta$-dependency 
(projection plus limb darkening effects; Equation (\ref{eq:flux})), 
while black dashed line shows the former effect alone.
\label{fig:flux_theta_dep}}
\end{figure}

Radiation flux ($F$) from 
 a unit surface area of an optically thick 
 disk toward a unit 
 solid angle at the polar angle of $\theta$ 
 decreases with an increasing $\theta $ as follows (Netzer 1987):
\begin{equation}
F(\theta) \propto \cos \theta \, (1 + 2 \cos \theta )
 \label{eq:flux}
\end{equation}
Here, the first term represents the change in the projected 
surface area, while the latter represents the limb darkening 
effect for plasma, whose opacity is dominated by 
electron scattering over absorption 
(Sunyaev \& Titarchuk 1985; Phillips \& Meszaros 1986).
Figure \ref{fig:flux_theta_dep} shows the $\theta$-dependency of $F(\theta)$, 
where the former effect alone is also drawn for comparison.
An accretion disk emits lesser 
radiation in the direction closer to its equatorial plane 
(i.e., larger $\theta $; 
Laor \& Netzer 1989; Sun \& Malkan 1989; Hubeny et al. 2000).
If the torus and the disk are aligned, 
the assumption of isotropic emission from accretion disks 
(e.g., Equation (\ref{eq:bar87})) obviously overestimates 
the radiation flux toward the torus,  
leading the overestimation of 
the inner radius of the torus. 

This effect works even if the disk 
is infinitesimally thin.
As shown in Section \ref{sec:qmax}, 
a nonzero thickness of the disk brings about another 
anisotropy of illumination flux, 
such that the torus is not 
illuminated below the disk height 
at $\theta $ larger than a critical angle $\theta_{\rm max} $.
Except in Section  \ref{sec:qmax}, 
we throughout adopt a thin disk with 
an aspect ratio of $\sim \! 0.01$, like 
the standard accretion disk model (Shakura \& Sunyaev 1973).

The effects of anisotropic emission and orientation 
 have been discussed  
 in the context of the Baldwin effect 
 in the line fluxes of photo-ionised emission  
(Netzer 1985; Francis 1993; Bottorff et al. 1997).
These effects upon the torus were examined for the first time 
in Paper I.

\subsection{Inner Structure of Torus}

The inner edge of the torus is determined so that 
the temperature of a clump (at the irradiated surface) equals 
 $T_{\rm sub}$ 
there. 
Since the radiation flux from the 
disk $F$ varies with the polar angle $\theta $, 
the sublimation
radius of the torus $R_{\rm sub}(\theta )$ is also a function of 
$\theta $.  
Namely, 
$R_{\rm sub}(\theta )$ 
is the distance between the torus edge and the central BH 
for various $\theta $.
In contrast, we express the sublimation 
radius estimated under the isotropic emission assumption 
(Equation (\ref{eq:bar87})) as $R_{\rm sub,0}$.
The anisotropic illumination given in Equation (\ref{eq:flux}) 
results in
\begin{equation}
R_{\rm sub}(\theta ) = R_{\rm sub,0} 
\left[\dfrac{\cos \theta \, (1 + 2 \cos \theta )}
{\cos \theta_{\rm obs} \, (1 + 2 \cos \theta_{\rm obs} )}\right]^{0.5}.
\label{eq:rsub}
\end{equation}
Here, $\theta_{\rm obs}$ is the polar angle toward the 
observer seen from the central accretion disk. 
Outside this radius, 
there are numerous clumps with their temperature 
below $T_{\rm sub}$. 
In the case of an isotropic emission from the disk, 
the torus edge is supposed to stand at a distance of $R_{\rm sub,0}$.
Although various grain sizes result in the sublimation 
process occurring over a transition zone rather than a single 
distance (Nenkova et al. 2008), we employ a sharp boundary 
for simplicity.

\begin{figure}
\plotone{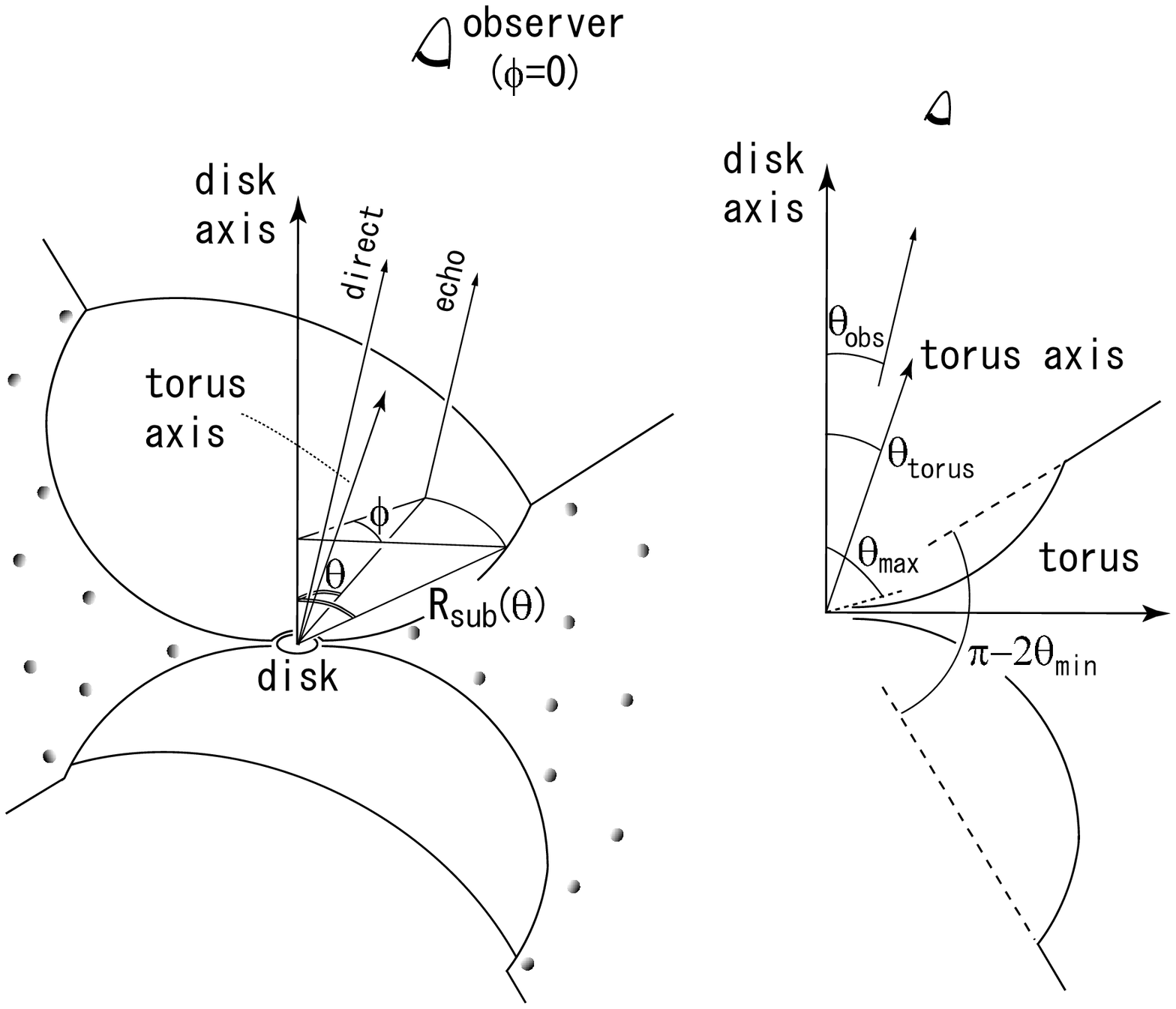}
\caption{Schematic view of the inner structure of the torus, 
for a misaligned configuration (Section \ref{sec:misaligned})
 with $\theta_{\rm torus} \sim 20$\degr\ and $\phi _{\rm torus} =0$\degr.\  
Right cartoon indicates the definitions of 
$\theta_{\rm min}$ (the torus thickness) and 
$\theta_{\rm max}$ (the disk thickness).
\label{fig:schematic}}
\end{figure}

In Paper I, we found that 
(1) the torus inner edge is located closer to 
the central BH than suggested by previous estimations 
(Equation (\ref{eq:bar87}))
and that (2) the structure of the edge is concave/hollow.
Moreover, (3) our result indicated that 
the innermost edge  of the torus may connect with the outermost
edge of the accretion disk continuously 
(e.g., Emmering et al. 1992; 
Elitzur \& Shlosman 2006). 
Figure~\ref{fig:schematic} shows a schematic view of the 
torus inner region.
If the torus is indeed a reservoir of gas for the disk, 
angular momenta of the infalling gas will 
align these axes.
The misalignment between the torus and the disk is investigated 
in Section \ref{sec:misaligned}.

\subsection{Calculation of Transfer Function} \label{sec:tf}

Current interferometric NIR observations (e.g., Swain et al. 2003; 
Kishimoto et al. 2009) and future Adaptive Optics imaging 
with $\sim \! 30$\,m telescopes cannot spatially resolve 
the innermost region (radius and shape) of the torus in nearby Seyfert galaxies.
Thus, observations of time variability will continue to be 
powerful tools to probe the innermost structure of the torus 
even in the coming decade.
We calculate the 
time variation of NIR emission in response 
to a $\delta$-function like variation of the 
optical/UV illumination [transfer function $\Psi (t)$]. 
Transfer functions contain various information of 
the re-emitting region, such as 
the shape and the emissivity profile 
etc (Blandford \& McKee 1982; Netzer 1990; Gaskell et al. 2007).
Time variation of the reprocessed radiation (NIR in this study) 
is a convolution of the illumination flux variation with 
$\Psi (t)$.
We calculate $\Psi (t)$ and its centroid $t_{\rm delay}$, 
which corresponds to the observed time lag.
Since the time profile is also one of the 
characteristics of this study compared to the earlier work 
(Sections  \ref{sec:phi_contri} and  \ref{sec:theta_contri}), 
we also calculate the width 
$rms$ and the skewness $s$ to describe the shape 
of $\Psi (t)$:
\begin{eqnarray}
rms &=& \left[ 
\int (t-t_{\rm delay})^2 \, \Psi (t) \, dt \,
/ \int \Psi (t) \, dt
  \right]^{0.5}\\
s &=& 
\int (t-t_{\rm delay})^3 \, \Psi (t) \, dt \,
/ \int \Psi (t) \, dt \, /\, rms^3  
\end{eqnarray}
A negative (or positive) $s$ means that the distribution 
has a 
left (or right) tail.
The computed $rms$ and $s$ would be useful to interpret and 
predict the cross correlation functions between NIR and 
optical/UV light curves. 
A larger $rms$ will correspond to a larger uncertainty in the 
measurements of the time delay.
As to the detectability of NIR variations, 
a small $rms$ and a high $\Psi (t)$ indicate the relative (i.e. in mag) 
and absolute (e.g., in erg\,s$^{-1}$) variability, respectively.

When the illumination flux varies, the inner edge of the torus
shifts in principle (Laor 2004).
Depending on whether dust grains in the clumps are sublimated or not, 
clumps belong to either 
the broad emission line region or the dusty torus
(Netzer \& Laor 1993; Suganuma et al. 2006).
However, it takes $\sim \! 1$\,year for the inner region 
of the 
torus to adjust to the varying illumination 
flux (Koshida et al. 2009; Pott et al. 2010). 
Thus, we regard that the inner structure of the dusty torus is steady 
in the timescale of NIR-to-optical time lag ($\sim$months).

To calculate $\Psi (t)$ for the clumpy torus, we consider the following 
items;
(1) the optical path, 
(2) NIR emissivity of the torus inner region as a function of $\theta $
and (3) anisotropic emission of each clump.
In this work, (4) 
we include the effect of torus 
self-occultation (i.e., absorption of NIR emission from a clump 
by other clumps on the line of sight). 
While the torus self-occultation is 
a minor effect for a typical type-1 AGN, 
it plays a significant role 
for 
inclined viewing angles, thick tori 
and misaligned tori.
Considering the self-occultation, 
we ignore the response from the aligned torus 
at $\theta > \frac{\pi }{2}$ (Appendix). 

First, (1) the optical path difference is written as
\begin{equation}
R_{\rm sub}(\theta ) 
\left[
1 - \{\cos \theta_{\rm obs} \, \cos \theta
 + \sin \theta_{\rm obs} \, \sin \theta \, \cos \phi \}
\right], 
\label{eq:opd}
\end{equation}
where $\phi$ is the azimuthal angle and defined so that 
$\phi = 0$ for the observer (Figure~\ref{fig:schematic}).
The concave shape of the inner region of the torus reduces the 
optical path difference.
Clumps at slightly farther 
and those at slightly closer than $R_{\rm sub}(\theta)$ 
will also emit NIR radiation,
smearing out the resultant NIR response. 
Since this effect 
unlikely changes the 
$\Psi (t)$ drastically, 
 we 
 consider only optical paths that hit the inner edge of the torus.

Next, (2) for the emissivity 
as a function of $\theta $,  
we assume that 
(2-1) the clump size increases and 
(2-2) the clump number density decreases 
when the clump-to-BH distance increases 
(e.g., H\"{o}nig et al. 2006; Schartmann et al. 2008).
Following Paper I,  
the emissivity of NIR flux per $d \Omega $ is 
assumed to be proportional to $R_{\rm sub}(\theta)^2$.

Third, (3) 
 the anisotropy of the NIR emission from each clump is considered,
 since clumps are optically thick to NIR (and optical/UV) 
radiation (see Appendix for details).
Namely, the question is  
 how extent 
the illuminated surface of a clump is seen by the observer 
(Nenkova et al. 2002).
Let us suppose that an observer looks at a clump with 
an angle $\xi $, where $\xi =0$ means a face-on view of the illuminated 
surface. 
We adopt the following anisotropic coefficient for 
the waning effect, 
\begin{equation}
\min \left[
1, \left(\frac{1 + \cos \xi}{2} + 0.1 \right)
\right].
\label{eq:waning}
\end{equation}
This coefficient is chosen so as to reproduce the Monte Carlo calculations 
by H\"{o}nig et al. (2006) for a single clump observed from 
three different $\xi $.

Finally, (4) if the line of sight from a region to the observer passes 
through the torus, 
we omit the NIR flux from such a region.
Radiation energy absorbed by the clumps on the way will be re-radiated 
at Mid-IR bands.

\begin{figure}
\plotone{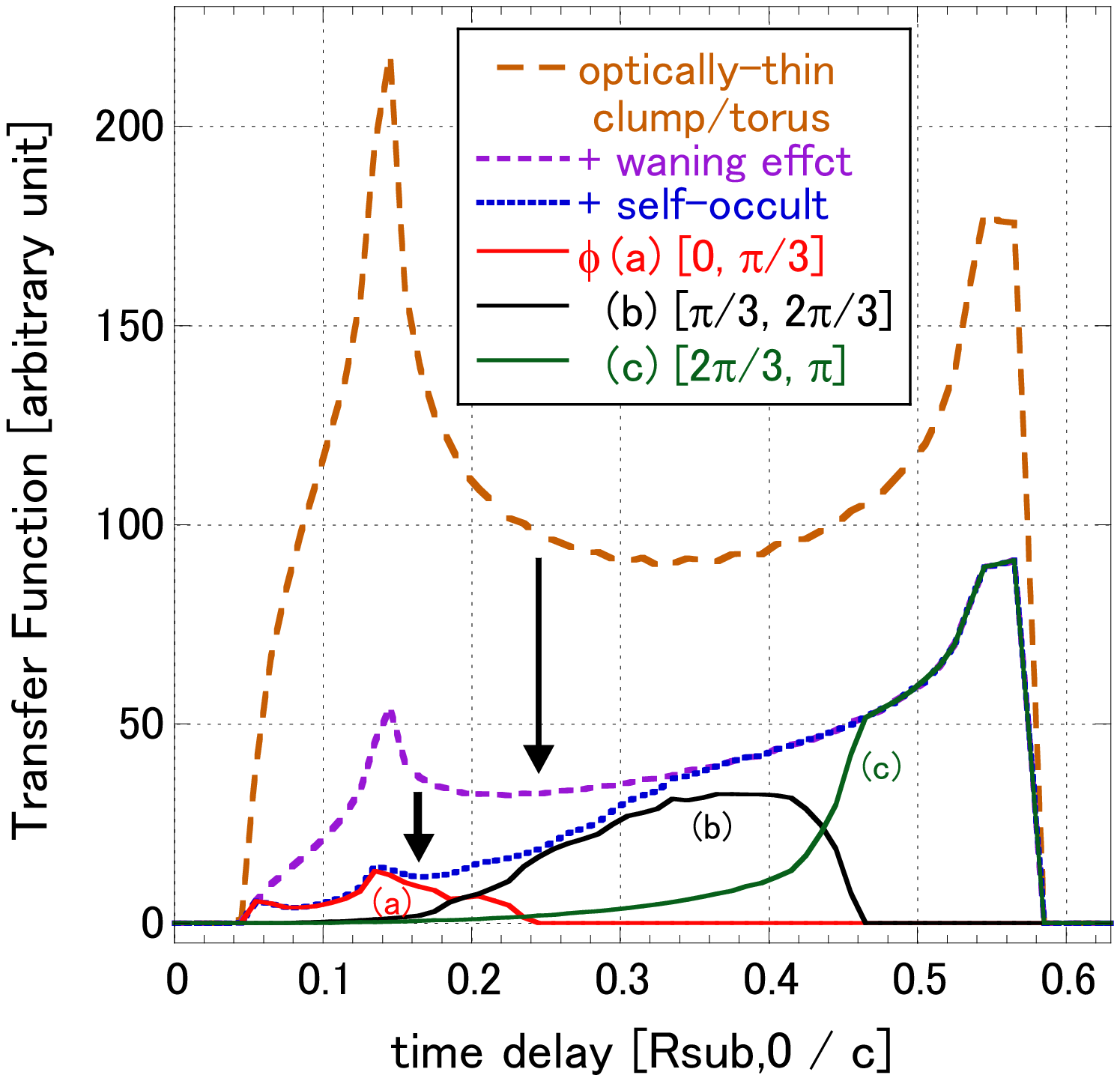}
\caption{
NIR response from $\theta < \frac{\pi}{2}$ of the aligned torus, 
viewed from $\theta_{\rm obs} = 25$\degr\ 
with $\theta_{\rm min} = 45$\degr\ and $\theta_{\rm max} = 89$\degr.
Brown long-dashed line 
is computed by switching off both the waning effect and the 
torus self-occultation, making 
the left horn higher than the right one similar to the 
$\Psi (t)$ presented by Barvainis (1992).
Purple short-dashed line then takes into account that 
each clump is optically thick, 
and is identical to the $\Psi (t)$ presented in Paper I.
Blue dotted line include both effects. 
Three solid lines show the contributions 
from each $\phi $ range.
The portion at (a) $\phi \sim 0$ produces the red left bump, 
which is  a faster 
(due to a shorter optical path difference)
and weaker (due to self-occultation and a stronger waning effect) response than 
the opposite area at (c) $\phi \sim \pi$ (green right horn).
\label{fig:phi_contribution}}
\end{figure}

The dotted line in Figure \ref{fig:phi_contribution}
 presents the resultant transfer function 
of the aligned torus 
 for $\theta_{\rm obs}$ = 25\degr, 
calculated by summing up the NIR responses from different 
portions of the torus at ($\theta$, $\phi$).
The integration is done from $\theta_{\rm min}$ 
to $\theta_{\rm max}$ in the $\theta$-direction and  0 to $\pi$
in the $\phi$-direction.
Here, the opening angle of the torus $\theta_{\rm min}$ 
is assumed to be 45\degr, 
which is roughly consistent with various 
 observational results (e.g., the ratio between types-1 and -2 AGNs and/or
 opening angles of light cones in the narrow line region), and 
the maximum $\theta$ of the torus 
$\theta_{\rm max}$ is set to 89\degr\ (i.e., thin disk approximation).
This parameter set is identical to the one 
adopted in Paper I, 
and is regarded as fiducial in this study.
The mean delay $t_{\rm delay}$, the width $rms$ and 
the skewness $s$ of $\Psi (t)$ are 
$0.42 \,R_{\rm sub,0}/c$, $0.13 \,R_{\rm sub,0}/c$ and $-0.76$, respectively.
For typical Seyfert 1 galaxies and quasars (e.g., with UV luminosity of 
$10^{43.5}$\,erg\,s$^{-1}$ and $10^{45.5}$\,erg\,s$^{-1}$), 
the expected inner radii $R_{\rm sub,0}$ are $\sim$\,0.1 
and 1\,pc (Equation \ref{eq:bar87}), while our resultant 
delays $t_{\rm delay}$ for the fiducial parameter set are about a month 
and a year, respectively (see Figure \ref{fig:tl}).

The long-dashed line 
is computed by switching off both the waning effect and the 
torus self-occultation, making 
the left horn higher than the right one similar to the 
time response of optically thin tori (Barvainis 1992).
Then, the short-dashed line takes into account the waning effect,
and is identical to the $\Psi (t)$ presented in Paper I, 
with $t_{\rm delay}$, $rms$ and $s$ 
of $0.37 \,R_{\rm sub,0}/c$, 
$0.15 \,R_{\rm sub,0}/c$ and $-0.35$, respectively.
The torus self-occultation results in a slight increase of $t_{\rm delay}$
and a more skewed profile.
We will see the reason for these changes in the next subsection.

\subsection{Response from various $\phi$} \label{sec:phi_contri}

In order to clarify contributions from different $\phi $, we divide 
the torus inner edge into three regions equally.
The three solid lines in Figure \ref{fig:phi_contribution}
show the $\Psi (t)$ from each $\phi $ range.
Among them, the left one is produced by the low $\phi $ region 
($\phi = 0 - \frac{\pi}{3}$), showing the rapid response 
at $t_{\rm delay} \sim 0.15 \, R_{\rm sub,0}/c$ 
(due to a short optical path difference) 
and a strong reduction of flux by the waning effect (large $\xi $) 
and the torus self-occultation.
On the contrary, the right one with the longer delay 
at $t_{\rm delay} \sim 0.5 \, R_{\rm sub,0}/c$ 
 comes from the 
large $\phi $ region ($\phi = \frac{2 \pi}{3} - \pi$), where 
clumps tend to direct their illuminated surface toward the 
observer, suffering from less waning effect.
We see that the torus self-occultation affects no influence at 
$\phi \ga \frac{\pi}{2}$ with the fiducial parameter set. 
In other words, both the waning effect and the self-occultation 
selectively reduce the emission from  the region with a short delay, 
making the right horn higher than the left one, 
opposing to the results for optically thin tori (Barvainis 1992). 
The fluences, $\int \Psi (t)\, dt$, from each $\phi $ region 
are 1.3, 6.4 and 10.4 from low to high $\phi $, respectively.

\subsection{Response from various $\theta$} \label{sec:theta_contri}

\begin{figure}
\epsscale{0.9}
\plotone{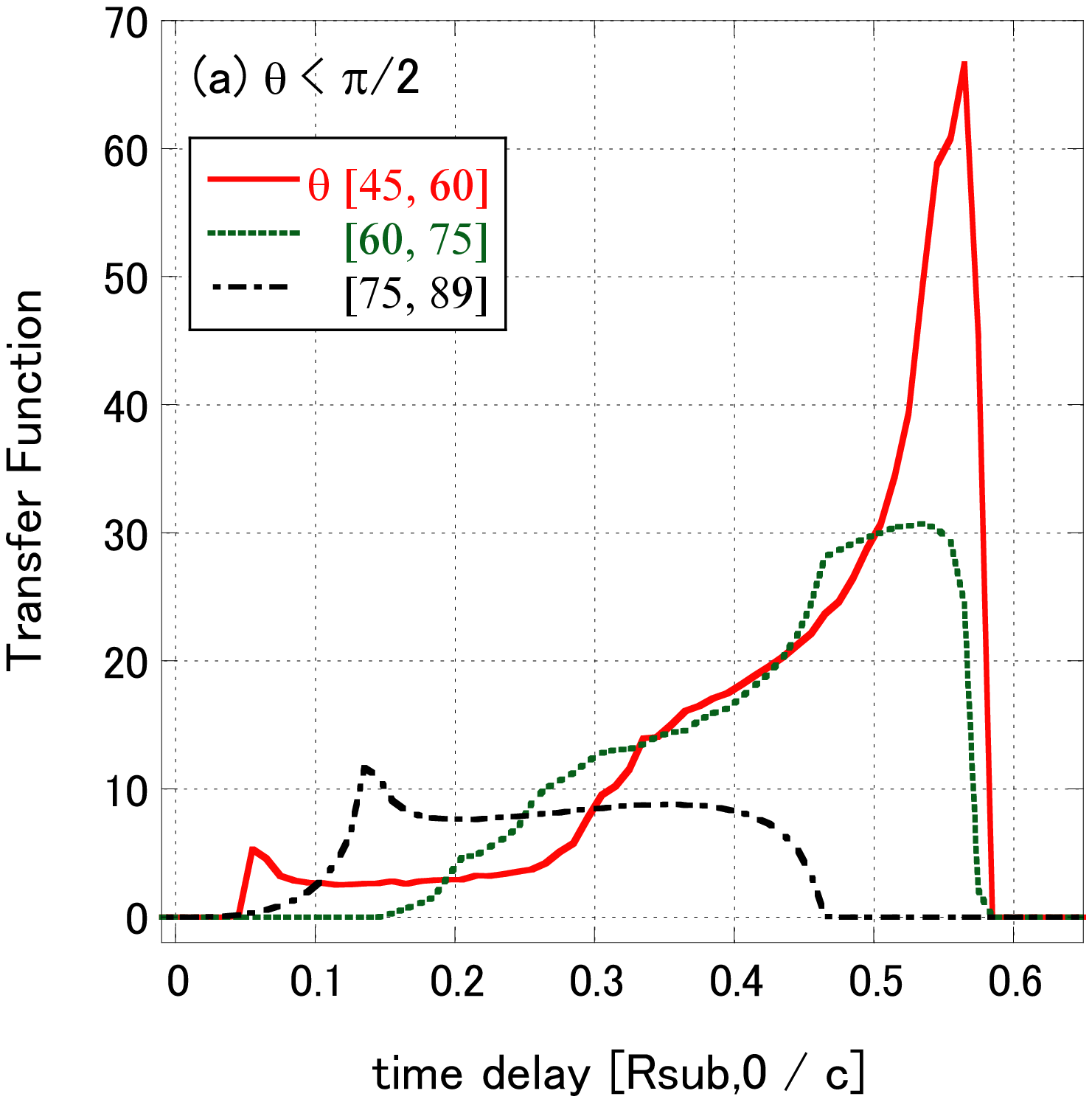}
\plotone{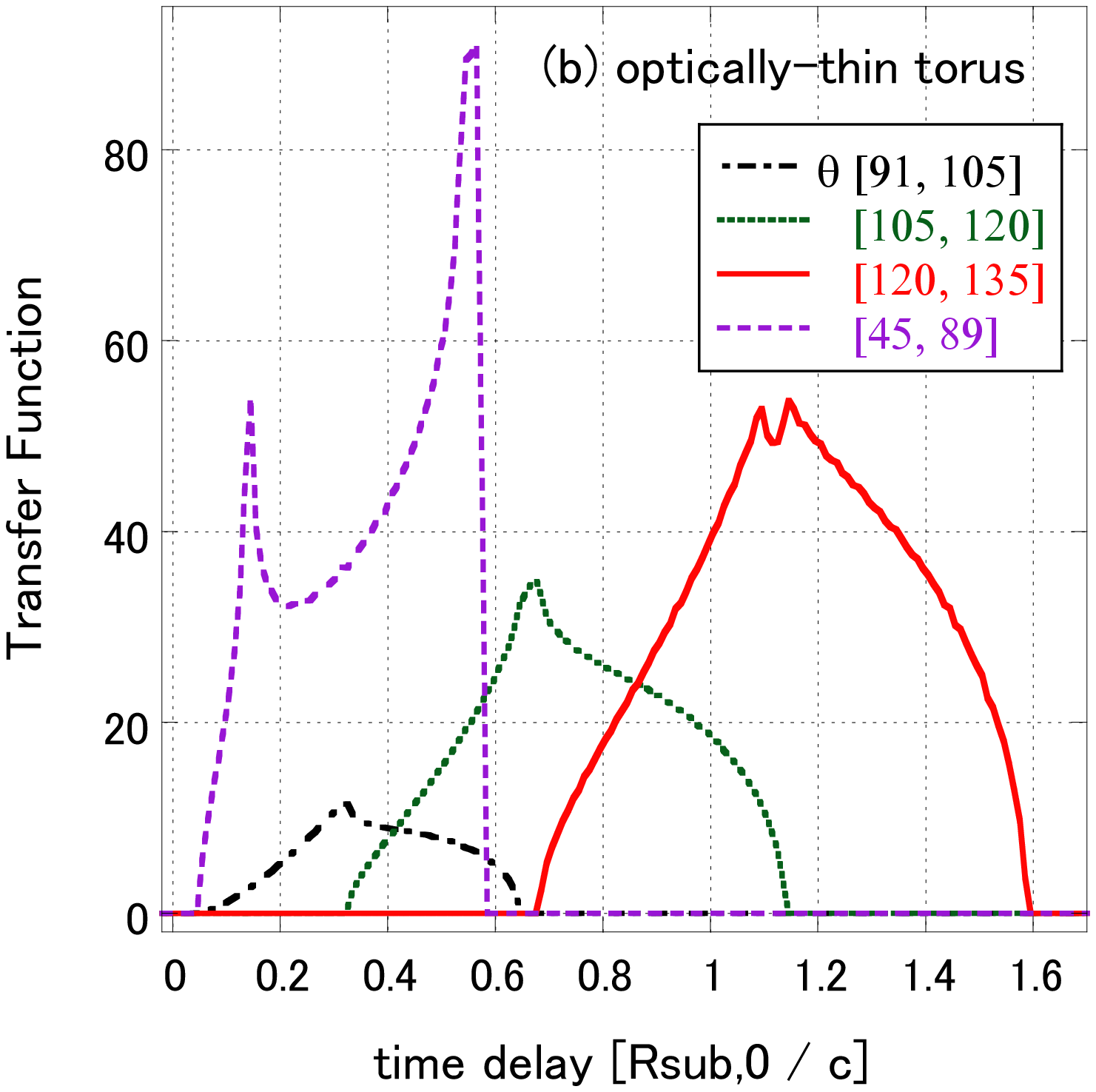}
\caption{
(a) Transfer functions from different $\theta$ ranges 
with the geometry same as Figure \ref{fig:phi_contribution};
45\degr--60\degr\ (red solid line), 
60\degr--75\degr\ (green dotted line), 
and 75\degr--89\degr\ (black dot-dashed line).
The NIR responses from $45\degr \le \theta \le 75$\degr\ 
appear at similar time delay.
(b) Transfer functions from the far side of the torus; 
$\theta=$\,91\degr--105\degr\ (black dot-dashed line), 
105\degr--120\degr\ (green dotted line), 
120\degr--135\degr\ (red solid line), computed without  
the torus self-occultation.
For comparison, $\Psi (t)$ from 45\degr--89\degr\ (purple short-dashed 
line, as in Figure \ref{fig:phi_contribution}) 
is also drawn.
If NIR emission from $\theta > \frac{\pi}{2}$ penetrates 
the torus and arrives at the observer, 
$t_{\rm delay}$ will be longer.
\label{fig:theta_contribution}}
\end{figure}

Later, we change the torus thickness (Section \ref{sec:qmin})
 and the disk thickness (Section \ref{sec:qmax}).
To understand how they will affect the time response of 
NIR emission, we draw $\Psi (t)$ from different $\theta $ ranges 
(at $\theta < \frac{\pi}{2}$) separately 
in Figure \ref{fig:theta_contribution}a.
It turns out that most NIR flux arises from  
small $\theta $ regions at $45\degr \le \theta \le 75$\degr\ 
(i.e., the region with a high latitude from the 
equatorial plane), which show similar time delay.
Because the emissivity is assumed in proportion to $R_{\rm sub}(\theta)^2$, 
the fluence from the largest $\theta $ range (75\degr--89\degr)
near the equatorial plane, where the torus inner edge is closest 
to the central BH, is small. 
The NIR fluences from each $\theta $ range,
from small to large $\theta $,  
is 8.4, 6.8 and 2.9, respectively.
Therefore, 
little 
difference in $t_{\rm delay}$ 
for various $\theta_{\rm max}$ and 
a shorter $t_{\rm delay}$ for an 
extremely thin ($\theta_{\rm min} \ga 75$\degr) torus
are expected.
Consequences of a thick torus are not drawn straightforwardly, 
because of 
the torus self-occultation. 

In case the torus is  
optically thin (cf. Appendix), 
we shortly mention the contribution from the far side 
($\theta > \frac{\pi}{2}$)
of the torus.
Depending on the origin of optical/UV time variability of 
AGNs (X-ray reprocessing 
or change in the accretion rate 
etc.; e.g., Kawaguchi et al. 1998; Sakata et al. 2011), 
the link between the disk fluxes 
to $\theta > \frac{\pi}{2}$ and 
to $\theta < \frac{\pi}{2}$ will vary.
The NIR time variations from $\theta > \frac{\pi}{2}$ 
is controlled by the variations of the disk flux to 
$\theta > \frac{\pi}{2}$, which we cannot observe.
Thus, if the disk illumination to the two sides are random, 
the NIR flux variations 
from $\theta > \frac{\pi}{2}$ 
influence the measurements of the optical/UV-to-NIR lag 
as noise.
On the other hand, if the time variations 
of disk illumination 
toward the two sides 
are similar, the time delay of NIR emission will become 
longer, as follows.
To calculate the NIR response from $\theta > \frac{\pi}{2}$, we 
replace $\cos \theta$ in Equation (\ref{eq:rsub}) by $|\cos \theta|$, 
and switched off the self-occultation effect.
Figure \ref{fig:theta_contribution}b 
 presents NIR responses from $\theta > \frac{\pi}{2}$, 
 which have long time delay due to their long light path.
If no extinction affects the NIR emission from 
$\theta > \frac{\pi}{2}$ (as Barvainis 1992 assumed), 
we will obtain the net response from $45\degr \le \theta \le 135$\degr\ 
with $t_{\rm delay}$ of $0.78 \,R_{\rm sub,0}/c$.
In other words, a long $t_{\rm delay}$ may be a signature of 
an extremely low volume filling factor of clumps in the torus 
and/or a very thin torus (Appendix).

\section{Aligned Torus: Various Dependencies} \label{sec:aligned}

In this section, 
we present various dependencies of the NIR emission
from the torus whose rotation axis is aligned to the disk axis.
At the end of each subsection, 
we shortly summarise the obtained dependency by contrasting 
with the result for the fiducial parameter set.

\subsection{Viewing Angle: $\theta_{\rm obs}$} \label{sec:qobs}

\begin{figure}
\epsscale{0.95}
\plotone{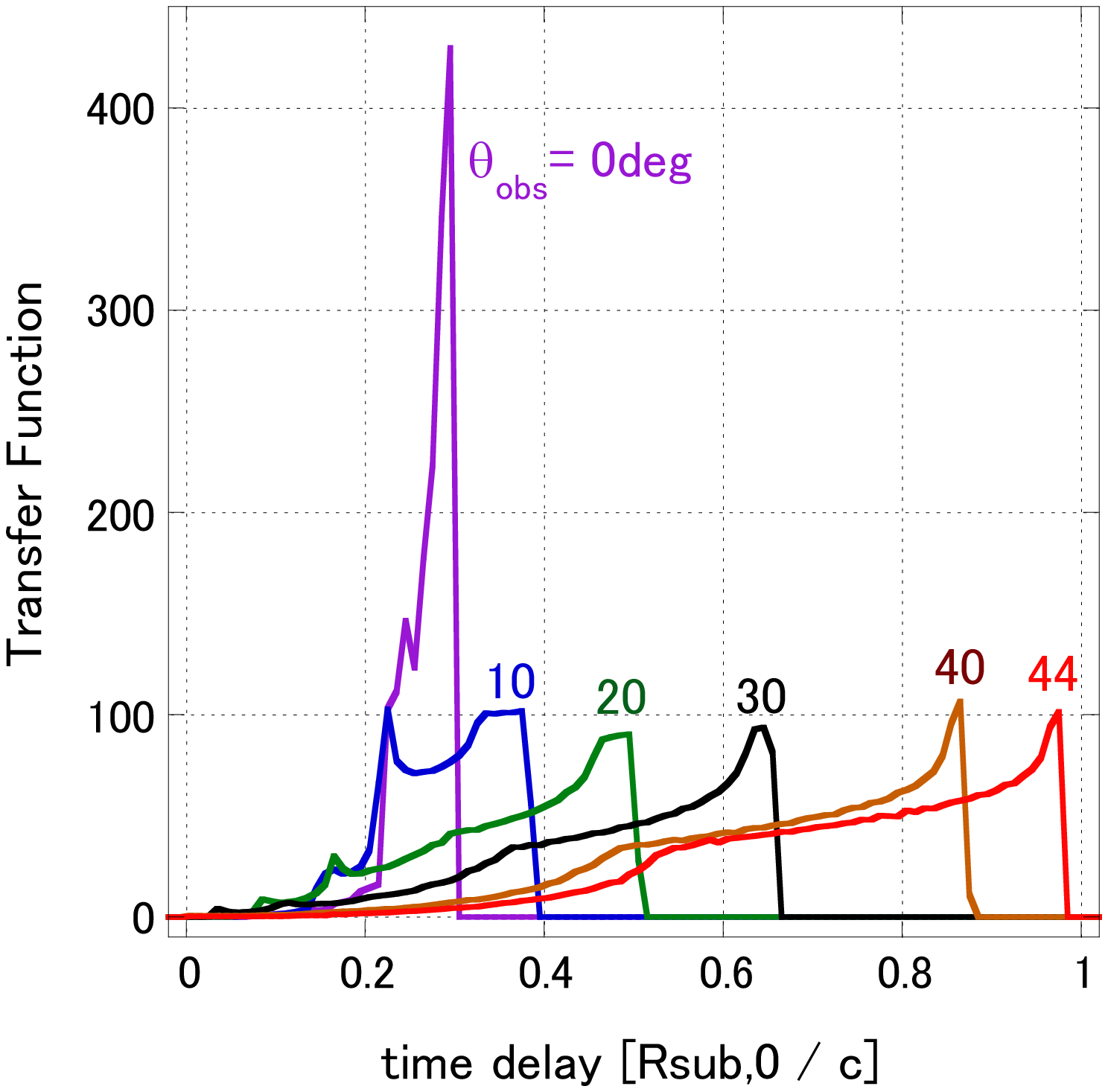}
\caption{
Transfer functions for various 
viewing angles, with 
$\theta_{\rm obs}$ labelled near each line,  
 from 0\degr\ (exactly pole-on geometry) 
 to 44\degr\ (presumably corresponding to type-1.9 AGNs).
Here, $\theta_{\rm min}$ and $\theta_{\rm max}$ are 
fixed at 45\degr\ and 89\degr, respectively.
As $\theta_{\rm obs}$ increases, 
(1) the centroid of the $\Psi (t)$ $t_{\rm delay}$ increases, 
(2) NIR flux [$\int \Psi (t)\, dt$] also increases, and 
(3) $\Psi (t)$ becomes broad. 
(4) For nearly face-on geometry, $\Psi (t)$ is peaky.
\label{fig:thetaobs_tf}}
\end{figure}

Figure \ref{fig:thetaobs_tf} shows the transfer functions for various 
$\theta_{\rm obs}$, from an exactly pole-on geometry 
($\theta_{\rm obs}=0$\degr) to 
inclined viewing angles.
With a large $\theta_{\rm obs}$ ($\approx 40\degr$--$44\degr$), 
the line of sight grazes the upper boundary of the torus, 
which would corresponds to the situation in type-1.8/1.9 AGNs.
Here, $\theta_{\rm min}$ and $\theta_{\rm max}$ are 
fixed at 45\degr\ and 89\degr, respectively.
As $\theta_{\rm obs}$ increases, we see that 
(1) the centroid of the response 
 $t_{\rm delay}$ increases, 
(2) the NIR fluence $\int \Psi (t)\, dt$ also increases, and 
(3) the profile 
becomes broad. 
In addition, 
(4) $\Psi (t)$ is quite peaky for a nearly face-on geometry.
The first and second results are our new findings. 
Although the third and forth trends are already reported 
for optically thin tori by Barvainis (1992), 
we find here that both trends are also true for optically thick tori.

\begin{figure}
\plotone{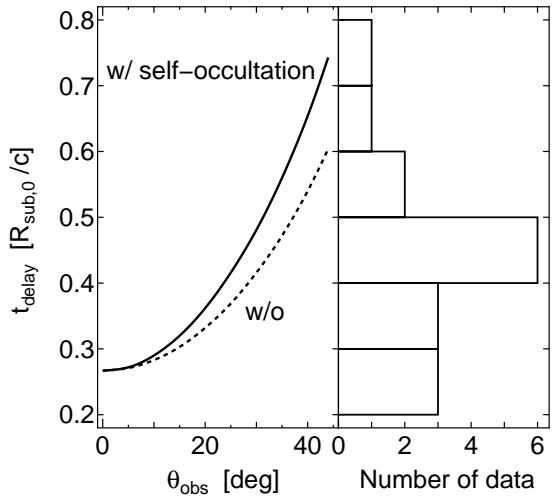}
\caption{
Centroid of $\Psi (t)$ $t_{\rm delay}$ as a function of 
$\theta_{\rm obs}$ (left). 
Short-dashed line is computed without the torus self-occultation.
While with this effect (solid line), 
the computed $t_{\rm delay}$ ranges from 0.27 to 0.74 
$R_{\rm sub,0}/c$. 
It covers the observed range of $t_{\rm delay}$ 
shown in the right histogram based on Figure \ref{fig:tl}, 
where 
the typical error of $t_{\rm delay}$ is $\sim 0.09 \, R_{\rm sub,0}/c$.
The viewing angle $\theta_{\rm obs}$ can be the primary parameter to 
cause the observed scatter about 
the regression line in the $t_{\rm delay}$--$L_{\rm UV}$ diagram. 
\label{fig:thetaobs_tdelay}}
\end{figure}

\begin{figure}
\plotone{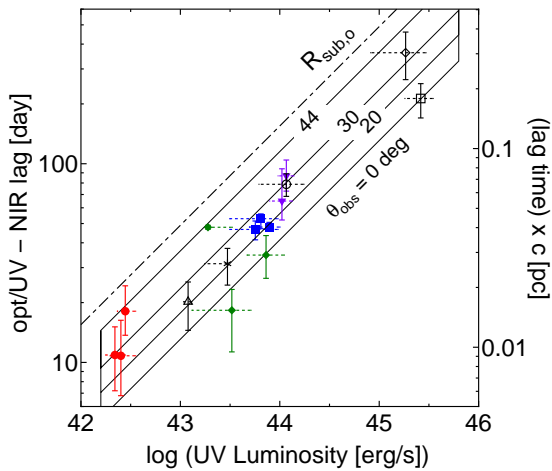}
\caption{
Relation between $t_{\rm delay}$ and the UV luminosity.
Corresponding scales are shown on the right axis.
Observed data are taken from Suganuma et al. (2006) 
and references therein, as done by Kishimoto et al. (2007). 
Objects with multiple data are plotted by filled symbols; 
NGC5548 ({\it blue squares}), NGC4051 ({\it red circles}), 
NGC7469 ({\it purple 
inverted triangles}) and NGC4151 ({\it green diamonds}).
Three data points for NGC4151 are collected in 1969-2001, 
and are indicative of a long-term evolution of the 
viewing angle.
Open symbols represent the objects with single data point for each: 
NGC3227 ({\it triangle}), Fairall 9 ({\it diamond}), 
GQ Com ({\it square}), NGC3783 ({\it circle}) and Mrk744 ({\it asterisk}).
Horizontal dashed lines show the ranges of the flux time variations.
Dot-dashed line represents $R_{\rm sub,0}$ (Equation \ref{eq:bar87}).
Loci for various $\theta_{\rm obs}$ (from $0\degr$ to $44\degr$) 
cover the observed scatter.
\label{fig:tl}}
\end{figure}

First, (1) $t_{\rm delay}$ is drawn as a function of 
$\theta_{\rm obs}$ in Figure \ref{fig:thetaobs_tdelay}.
For comparison, the result computed without the self-occultation 
is also shown.
The torus veils selectively the region with a short delay 
(Figure \ref{fig:phi_contribution}), hence enlarges $t_{\rm delay}$.
The self-occultation shows larger influences for more inclined angles.
On the right-hand side, we also draw a histogram of the 
observed delay in the unit of $R_{\rm sub,0}/c$ 
(
based on Figure \ref{fig:tl}).
The computed $t_{\rm delay}$ ranges from 
0.27 to 0.74\,$R_{\rm sub,0}/c$ 
(0.27 to 0.60\,$R_{\rm sub,0}/c$ without the self-occultation), 
which covers the range of the observed time delay.
In contrast, since Barvainis (1992) assumed an optically thin torus, 
$t_{\rm delay}$ was expected to have no (or quite weak) 
$\theta_{\rm obs}$-dependency.

Since such a broad range is not achieved by the changes of 
$\theta_{\rm min}$ and $\theta_{\rm max}$ as we will see 
later, we propose that the viewing angle is the key parameter 
responsible for the observed scatter about 
the regression line in the $t_{\rm delay}$--$L_{\rm UV}$ 
diagram (Oknyanskij \& Horne 2001; Suganuma et al. 2006).
Conversely, 
the measurements of $t_{\rm delay}$ 
are potentially useful 
to estimate the inclination angles.

In order to compare our results with the observed data 
more directly, 
the $t_{\rm delay}$ v.s. $L_{\rm UV}$ diagram 
is drawn in Figure \ref{fig:tl} (Suganuma et al. 2006 
and references therein).
Following Kishimoto et al. (2007), we estimate $L_{\rm UV}$ by 
$6 \, \nu L_\nu (V)$.
The uncertainty of the $t_{\rm delay}$ is $\sim 0.09 R_{\rm sub,0}/c$ 
on average.
Our loci for various $\theta_{\rm obs}$ 
(with $\theta_{\rm min}$ and $\theta_{\rm max}$ fixed at 
  the fiducial values)
well cover the observed scatter.
A type-1.5 Seyfert galaxy NGC3227 (triangle) is located at an area with 
a small $\theta_{\rm obs}$  ($0\degr$--$30\degr$), which would 
require 
a hysteresis effect (Koshida et al. 2009)
or a thin torus
for its relatively short $t_{\rm delay}$.
Among the three points for NGC4151 (green diamonds), 
the lower two data 
(collected in 1969-1980 and 1990-1998) are consistent with a pole-on 
view, while the upper one in 2001 indicates an inclined angle.
A change of the viewing angle on the timescale of 
tens years, due to e.g., a precession of the disk, is indicated.

\begin{figure}
\plotone{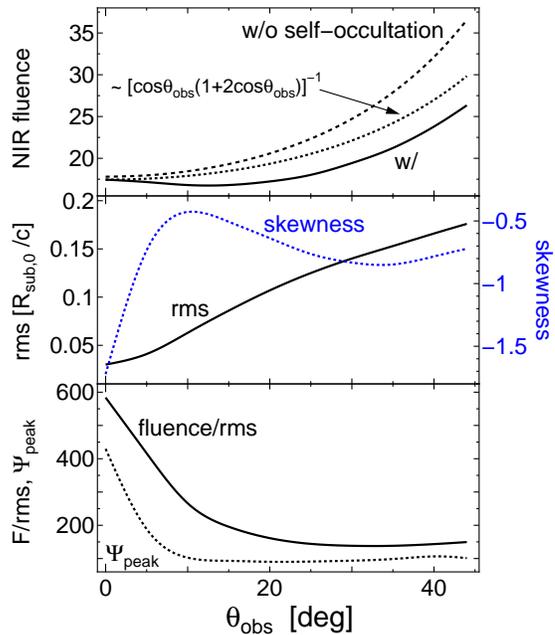}
\caption{
{\it Top}:
NIR fluence, $\int \Psi (t)\, dt$, as a function of 
$\theta_{\rm obs}$. 
As Figure \ref{fig:thetaobs_tdelay}, solid and short-dashed 
lines mean the results with and without the torus 
self-occultation.
Dashed line indicates the disk flux toward $\theta=0$\degr, 
normalised so that it coincides with the NIR fluence 
at $\theta_{\rm obs} = 0$\degr.
{\it Middle}:
Width of the transfer functions $rms$ (solid line) 
and the skewness $s$ (blue dotted line). 
Inclined viewing angles lead to broader $\Psi (t)$.
{\it Bottom}:
NIR fluence divided by $rms$ (solid line) and 
peak value of the transfer functions (dashed line), 
which describe the peakiness of $\Psi (t)$.
At a nearly pole-on view with $\theta_{\rm obs} \la 10$\degr, 
$\Psi (t)$ is peaky, implying the NIR variability is easily detected.
\label{fig:thetaobs_lssp}}
\end{figure}

Second, (2) 
the top panel of Figure \ref{fig:thetaobs_lssp} 
shows that the 
NIR fluence toward the observer, $\int \Psi (t)\, dt$, 
is insensitive to the viewing angle at $\theta_{\rm obs} \la 25\degr$ 
 and increases with  
$\theta_{\rm obs}$ at more inclined angles (solid line). 
Comparing $\theta_{\rm obs} = 0$\degr\ with $44$\degr, 
the fluence increases by a factor of 1.5. 
Due to the self-occultation, the NIR flux decreases 
by 19\% at $\theta_{\rm obs} = 25$\degr\ and 
by 26--28\% at $\theta_{\rm obs} = 40\degr$--$44\degr$.

Regarding a continuous illumination from the disk as 
a series of 
flash, 
we can draw conclusions other than the time response of 
NIR emission.
Namely, the obtained NIR fluence indicates the NIR flux toward the observer 
under a given steady optical/UV illumination.
We recall that we compute for various objects with 
a common 
optical flux toward the observer.
Therefore, the $y$-axis (NIR fluence) directly indicates the 
NIR-to-optical color.
In other words, the NIR-to-optical color becomes red as 
$\theta_{\rm obs}$ increases.
We thus argue that the red IR-optical color observed for type-1.8/1.9 
AGNs compared with type-1 objects (Alonso-Herrero et al. 2003) arises 
not only from dust extinction but also from their intrinsically red color.

The red color for large $\theta_{\rm obs}$ objects 
can originate in the reduction of optical flux toward 
large $\theta$ (Equation (\ref{eq:flux})) and/or in the change 
of NIR flux.
Here, we try to distinguish its origin.
The dotted line shows the disk flux toward $\theta=0$\degr, 
in proportion to 
$[\cos \theta_{\rm obs} \, (1 + 2 \cos \theta_{\rm obs})]^{-1}$, 
normalised so that it coincides with the NIR fluence 
at $\theta_{\rm obs} = 0$\degr.
If the NIR emission is isotropic, the NIR fluence in this diagram 
should change as the dotted line.
Instead, if the computed NIR fluence is smaller (or larger) than the 
dotted line, it means the NIR flux decreases (or increases) with 
$\theta$.
Our result indicates that the NIR flux decreases with 
$\theta$ due to the torus self-occultation, but its $\theta$-dependency 
is weaker than that of the disk flux, 
thereby presenting the red NIR-optical color for inclined angles.
The reduction of the NIR flux toward a larger $\theta_{\rm obs}$ 
is consistent with the radiative transfer calculations under 
an isotropic optical/UV illumination (H\"{o}nig et al. 2006; Nenkova et al. 2008).
Incorporation of the anisotropy of the optical/UV flux from the 
disk (Equation (\ref{eq:flux})) enables us to predict 
the NIR-optical/UV color.

Third, (3) 
the middle panel of Figure \ref{fig:thetaobs_lssp} 
describes the profile of $\Psi (t)$.
As $\theta_{\rm obs}$ increases, the width of 
$\Psi (t)$ $rms$ also increases, 
meaning that $\Psi (t)$ becomes broader.
Both $t_{\rm delay}$ and $rms$ increase with $\theta_{\rm obs}$, 
and their ratio $t_{\rm delay} / rms$ slightly rises toward 
a smaller viewing angle: 
a $t_{\rm delay} / rms$ ratio increases twice 
between $\theta_{\rm obs}=40$--$44\degr$
and $0\degr$.
With a small $\theta_{\rm obs}$, the echo from various 
parts of the inner edge arrives at the observer at 
a similar delay, making the $\Psi (t)$ quite narrow.
For larger $\theta_{\rm obs}$, a variety of light pass difference 
arise between $\phi \sim 0$ and $\sim \pi$, which causes 
the broader $\Psi (t)$.
We expect that the cross correlation function between 
NIR and optical/UV flux variations becomes broader 
for type-1.5--1.9 AGNs compared with a typical type-1 object.
The skewness $s$ is always negative (i.e., with a tail toward a shorter 
time delay), and the degree of asymmetry gets larger for a pole-on view.

Finally, (4) we comment on the NIR variability amplitude 
(to be precise, the ratio of the NIR amplitude to the optical/UV one).
A peaky $\Psi (t)$ will result in a large NIR amplitude, 
whereas a less peaky $\Psi (t)$ smears out the variability 
of reprocessed emission, 
producing a less NIR amplitude.
In order to see this quantitatively, we draw 
the NIR fluence-to-$rms$ ratio and the peak value of $\Psi (t)$ 
as a function of $\theta_{\rm obs}$ in 
the bottom panel of Figure \ref{fig:thetaobs_lssp}.
For nearly pole-on view with $\theta_{\rm obs} \lesssim 10$\degr,
both quantities rise.
For a given optical/UV variability, 
such pole-on objects will show large NIR variability amplitudes.
At $\theta_{\rm obs} \ga 15$\degr,
the peakiness of $\Psi (t)$ is insensitive to the viewing angle.

By contrast with the fiducial $\theta_{\rm obs}$ of 25\degr, 
objects with a small viewing angle will exhibit a short 
time delay 
with a narrow and peaky response.
On the other hand, a more inclined viewing angle leads to 
a longer delay with a broader profile and 
to an intrinsically redder NIR-to-optical color.
The $\theta_{\rm obs}$-dependent delay contrasts our work 
with the earlier study for optically thin tori (Barvainis 1992).
The computed range of $t_{\rm delay}$ coincides 
with the observed one.
The NIR response always shows an asymmetry 
with a tail toward a shorter delay.

\subsection{Torus Thickness: $\theta_{\rm min}$} \label{sec:qmin}

In Paper I, we assumed that 
the semi-thickness of the torus (from the equatorial plane 
to the upper surface) is 
45\degr.
However, luminous AGNs (quasars) seem to have thinner tori 
than Seyfert galaxies (Lawrence 1991; 
Ueda et al. 2003; 
La Franca et al. 2005; 
Arshakian 2005; Simpson 2005;
Maiolino et al. 2007;
Hasinger 2008;
Treister et al. 2008).
Moreover, recent hard X-ray observations discovered type-2 
AGNs with very thick tori (Levenson et al. 2002; 
Ueda et al. 2007; Eguchi et al. 2009; 
Noguchi et al. 2010).
In this subsection, we show the expected characteristics of the NIR 
emission from type-1 AGNs with thick and thin tori.

\begin{figure}
\epsscale{0.95}
\plotone{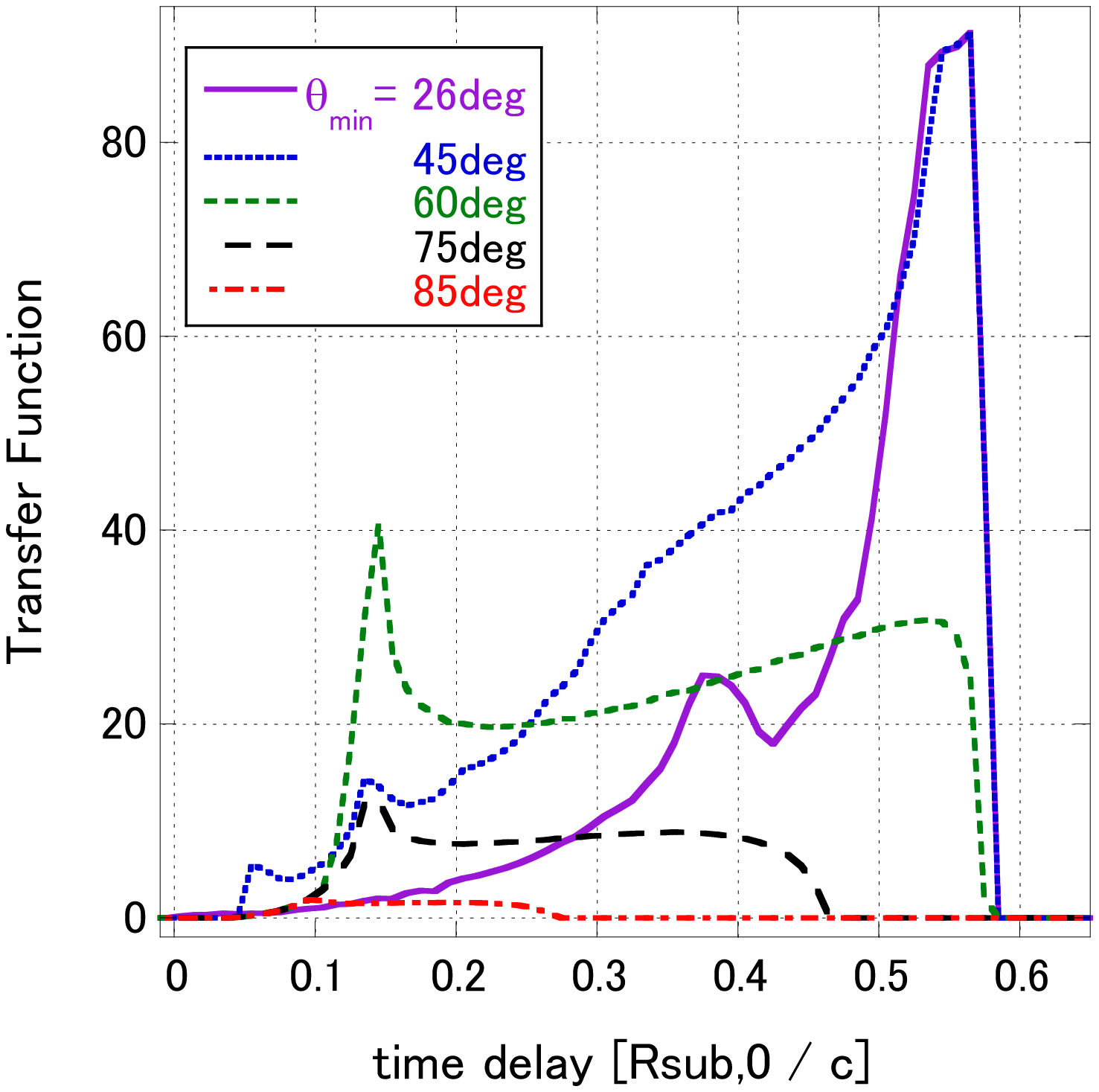}
\caption{
Transfer functions for various torus thickness $\theta_{\rm min}$.
A smaller $\theta_{\rm min}$ corresponds to a thicker torus.
The torus self-occultation 
by the $\theta \la 60\degr$ region at 
$\phi \sim 0$ hides the left horn, which originates in the large  
$\theta$ ($60\degr \la \theta \la 80\degr$) region at $\phi \sim 0$ 
and is visible in green short-dashed line.
\label{fig:thetamin_tf}}
\end{figure}

Figure \ref{fig:thetamin_tf} shows transfer functions for various 
$\theta_{\rm min}$, with $\theta_{\rm obs}$ and $\theta_{\rm max}$ 
fixed at 25\degr\ and 89\degr, respectively.
When the torus thickens from $\theta_{\rm min} = 45\degr$
to $26\degr$, 
the self-occultation selectively 
veils the region with a short delay 
at $\phi \sim 0$, 
hence enlarges the delay 
and reduces the fluence, the width 
and $s$. 
In other words, the self-occultation 
by the region with $\phi \sim 0$ 
and $\theta \la 60\degr$ hides the rapid response (left horn), 
which originates in the large 
$\theta$ ($60\degr \la \theta \la 80\degr$) region 
and becomes visible 
when the torus gets thin 
($\theta_{\rm min} = 60$\degr).
On the other hand, the self-occultation unlikely influences the 
result for a thin torus.
As shown in Section \ref{sec:theta_contri}, 
a rapid response is obtained when the torus 
is extremely thin ($\theta_{\rm min} \ga 75$\degr), 
since $R_{\rm sub} (\theta)$ is small at large $\theta$.

\begin{figure}
\plotone{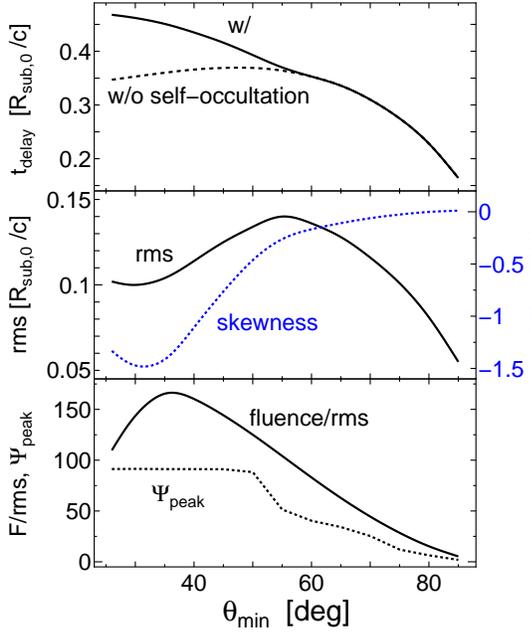}
\caption{
{\it Top}:
Time delay (in the unit of $R_{\rm sub,0}/c$) as a function 
of torus thickness $\theta_{\rm min}$.
The meanings of different lines are the same as Figures 
\ref{fig:thetaobs_tdelay} and \ref{fig:thetaobs_lssp}.
A thick torus  
veils the region with a short light path 
difference at $\phi \sim 0$, and hence enlarges $t_{\rm delay}$
and reduces $rms$ at $\theta_{\rm min} \la 55\degr$.
As the torus becomes thinner, 
$t_{\rm delay}$ becomes shorter.
{\it Middle} and {\it Bottom}: 
The same as Figure \ref{fig:thetaobs_lssp}, but for 
the $\theta_{\rm min}$-dependency here.
As the torus becomes thicker, 
$\Psi (t)$ shows a longer delay with a narrower and more heavily 
skewed profile due to the torus self-occultation.
A less peaky $\Psi (t)$ is expected for a thin torus, making 
the detectability of NIR variability difficult.
\label{fig:thetamin_tlss}}
\end{figure}

To see the $\theta_{\rm min}$-dependency more quantitatively, 
we draw $t_{\rm delay}$ as a function of $\theta_{\rm min}$ 
in the top panel of Figure \ref{fig:thetamin_tlss}.
Clearly, 
$t_{\rm delay}$ becomes short for thin tori.
Therefore, we expect that luminous quasars will show a relatively 
short time delay in the unit of $R_{\rm sub,0}/c$.
For instance, if the torus thickness is reduced for 
luminous objects, the loci in Figure \ref{fig:tl} shown by the 
solid lines become bent (convex).
The computed $t_{\rm delay}$ ranges from 0.16 to 0.47\,$R_{\rm sub,0}/c$
for $\theta_{\rm min}$ in $85\degr$--$26\degr$.
In contrast to the $\theta_{\rm obs}$-dependency, 
it seems difficult to explain the observed range 
(shown as the histogram in Figure \ref{fig:thetaobs_tdelay})
by a change of $\theta_{\rm min}$ alone.

As the torus becomes thicker, 
$\Psi (t)$ shows a longer delay 
with a narrower and more heavily skewed profile  
at $\theta_{\rm min} \la 55\degr$  
due to the torus self-occultation 
(middle panel). 
This trend opposes to the results for various viewing angles, where 
$t_{\rm delay}$ and $rms$ are positively correlated each other.
Although a low $\Psi (t)$ is expected for a thin torus (bottom panel), 
the small $rms$ implies 
a large relative variability in the NIR emission.

Next, Figure \ref{fig:thetamin_tf} also shows that the NIR 
fluence is a strong function of $\theta_{\rm min}$.
Qualitatively, it is trivial, since various $\theta_{\rm min}$ 
mean various solid angles of the torus subtended at the 
central BH, $\Omega_{\rm torus}$.
Here, 
\begin{equation}
\frac{\Omega_{\rm torus}}{4 \pi}
= \frac{ \int^{2 \pi}_{0} \int^{\theta_{\rm max}}_{\theta_{\rm min}} 
  \sin \theta \, d\theta \, d\phi }{2 \pi}
= \cos \theta_{\rm min} - \cos \theta_{\rm max}.
\label{eq:omega}
\end{equation}
However, as shown in numerical results by Nenkova et al. (2008), 
the NIR flux is not exactly in proportion to $\Omega_{\rm torus}$.
[Therefore, we need a caution when relating the observed NIR-to-UV 
luminosity ratio with $\Omega_{\rm torus}$ 
(e.g., Mor \& Trakhtenbrot 2011; see also Section \ref{sec:qmax}).]
Moreover, we let the inner radius of the torus 
$R_{\rm sub}(\theta)$ vary with $\theta$.
Thus, it is not obvious how the NIR fluence of our 
torus model varies with $\Omega_{\rm torus}$. 

\begin{figure}
\plotone{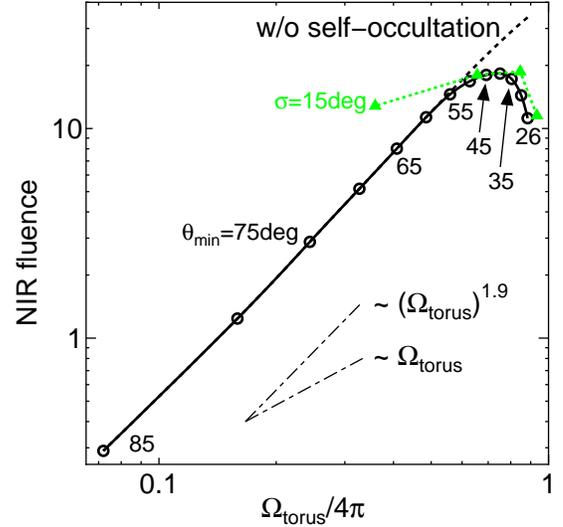}
\caption{
NIR fluence, $\int \Psi (t)\, dt$, 
to the observer at $\theta_{\rm obs} = 25\degr$ 
for various torus thickness $\theta_{\rm min}$, 
as a function of the solid angle of the torus 
seen from the central BH $\Omega_{\rm torus}$.
Solid and short-dashed lines have the same meanings as Figures 
\ref{fig:thetaobs_tdelay}, \ref{fig:thetaobs_lssp}  
and \ref{fig:thetamin_tlss} top panels.
For the former, circles are also plotted in $5\degr$ steps of 
$\theta_{\rm min}$, and labelled in $10\degr$ steps.
As the torus becomes thicker with 
$\theta_{\rm min} \la 40$\degr, the NIR flux starts to 
decreases owing to the torus self-occultation.
For a thin torus with $\theta_{\rm min} \ga 55$\degr, the NIR fluence is in proportion 
to $\Omega_{\rm torus}^{1.9}$.
Green dotted line with triangles is based on the 2$\mu$m flux densities 
calculated 
by Nenkova et al. (2008), for the torus semi-thickness parameter $\sigma $ 
of $15\degr$, $30\degr$, $45\degr$ and $60\degr$ 
(from the left to the right).
A modest thickness of the torus leads to the strongest NIR emission.
\label{fig:thetamin_fluence}}
\end{figure}

Figure \ref{fig:thetamin_fluence} 
shows that the NIR fluence 
decreases drastically as $\theta_{\rm min}$ increases.
A thin torus 
locates its inner 
radius at a short distance from the BH 
(Equation (\ref{eq:rsub}) and Figure \ref{fig:schematic}), 
where the size of clumps is small and the NIR emissivity is low 
(Section \ref{sec:tf}).
For thin tori with $\theta_{\rm min} \ga 55$\degr, 
we find that the NIR fluence is roughly in proportion 
to $\Omega_{\rm torus}^{1.9}$. 
Thus, luminous quasars are expected to show weak 
NIR emission (blue NIR-to-optical color), 
which is consistent with the observed trend of 
the decreasing NIR-to-optical flux ratio 
with an increasing optical luminosity 
(Maiolino et al. 2007; Treister et al. 2008; 
Jiang et al. 2010; Mor et al. 2011).
AGNs with very weak NIR emission 
(such as "hot-dust-poor" AGNs named by Hao et al. 2010) 
may indicate that their tori are very thin.
For comparison, we also draw the flux densities 
at 2\,$\mu$m from clumpy tori 
seen from $\theta_{\rm obs} = 25$\degr, 
based on a radiative transfer computation 
(Figure 8 of Nenkova et al. 2008, 
corrected following its Erratum by Nenkova et al. 2010, and
scaled to match with our result at 
$\theta_{\rm min} \sim 45$\degr).
They present the results for four different 
torus semi-thickness $\sigma$, 
with a gaussian clump distribution. 
For the smooth 
boundaries of the torus in the $\theta$ direction, 
the solid angle is computed as (see Apendix):
\begin{equation}
\frac{\Omega_{\rm torus}(\sigma)}{4 \pi}
= \int^{\frac{\pi}{2}}_{0} 
  \left( 1-\exp \left[-5 \, 
  \exp \left(-\frac{(\frac{\pi}{2}-\theta)^2}{\sigma^2} \right) \right] \right)
  \sin \theta \, d\theta.
\end{equation}
which is larger than $\cos \left( \frac{\pi}{2}-\sigma \right)$.
The steeper decline of the NIR flux in our result is likely 
due to the $\theta$ dependency of $R_{\rm subl} (\theta )$,
 in contrast to the constant sublimation radius (resulted from 
 the presumed istropic illumination) in their computations.

Next, we move on to the result for a thick torus.
The self-occultation reduces the NIR flux by 
55--67\% for $\theta_{\rm min} = 30\degr$--$26\degr$, 
whereas it results in a reduction by $19$\% for 
the fiducial $\theta_{\rm min}$ of $45\degr$.
Due to the self-occultation, too thick tori with 
$\theta_{\rm min} < 40$\degr\ exhibit lesser NIR flux 
(i.e., intrinsically bluer NIR-optical color)
than a torus 
with a moderate thickness with $\theta_{\rm min} \sim 40\degr$--$45\degr$.
The flux reduction of a thick torus compared with a modestly thick 
torus is again consistent 
with the result by Nenkova et al. (2008).

Bringing together the two behaviours, 
both a thick and a thin tori show the weak NIR emission.
Namely, 
a modest thickness of the torus 
leads to the strongest NIR emission. 
Therefore, a selection bias will arise such that NIR-selected 
AGNs tend to possess moderately thick tori.

Compared with the modest torus thickness ($\theta_{\rm min} = 45\degr$), 
both a thick and a thin tori display the weaker NIR emission 
(i.e., a bluer NIR-optical color).
A thick torus shows a slightly delayed, narrow and largely 
skewed NIR response.
On the other hand, 
 as the torus gets thinner, 
the NIR response becomes more rapid, narrower,  
closer to time-symmetric and low. 

\subsection{Disk Thickness: $\theta_{\rm max}$} \label{sec:qmax}

\begin{figure}
\epsscale{0.95}
\plotone{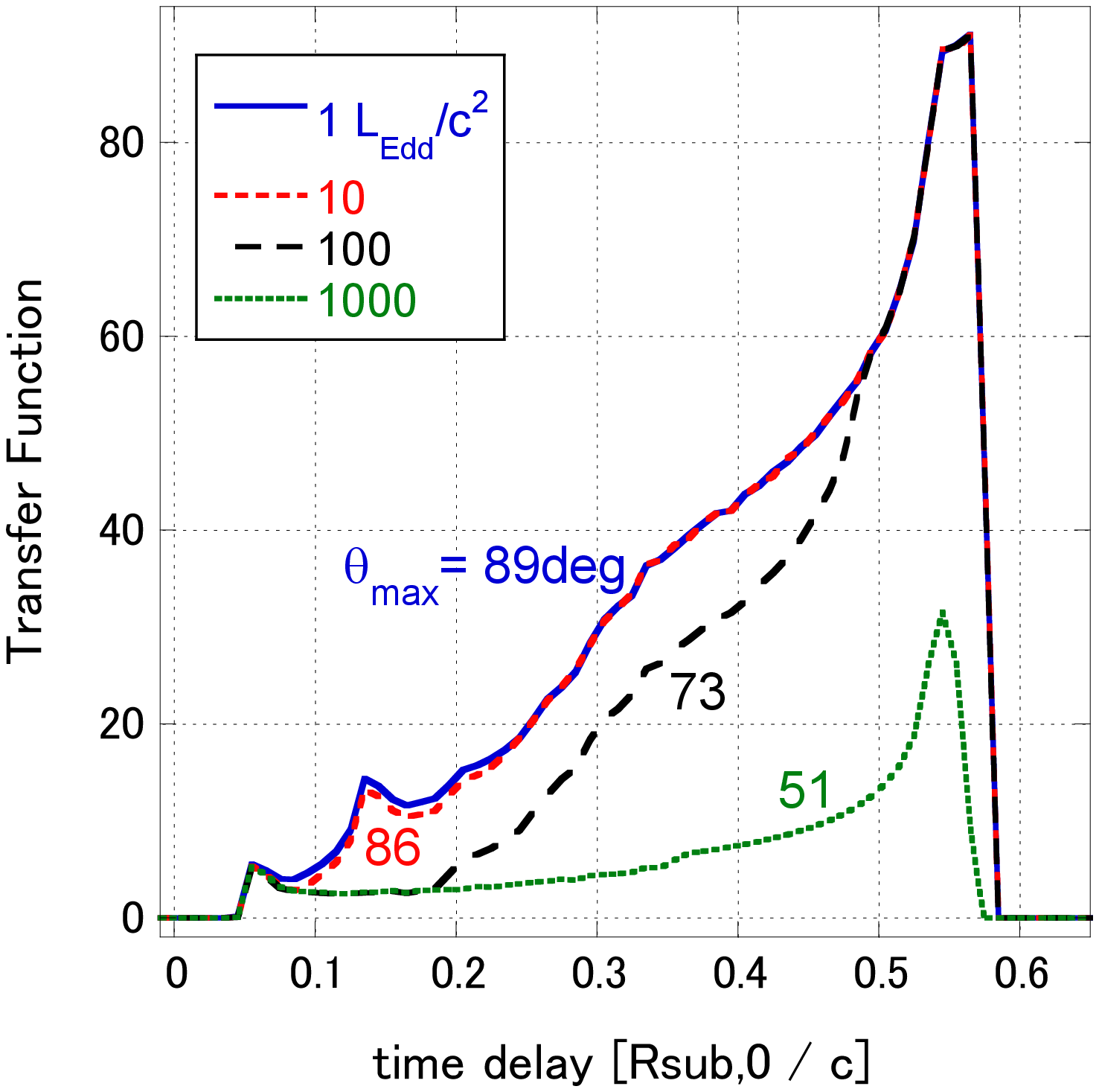}
\caption{
Transfer functions for various accretion rates, 
with $\theta_{\rm max}$ labelled near each line.
Here, $\theta_{\rm obs}$ of $25\degr$ and 
$\theta_{\rm min}$ of $45\degr$ are assumed.
As mentioned in Section \ref{sec:theta_contri}, 
$t_{\rm delay}$ seems insensitive 
to $\theta_{\rm max}$.
\label{fig:thetamax_tf}}
\end{figure}

When the accretion rate exceeds the Eddington rate
($\approx \! 16 L_{\rm Edd}/c^2$), 
an optically thick
advection-dominated accretion flow (a slim disk)
appears (Abramowicz et al. 1988). 
Since super-Eddington disks are geometrically thick 
(Abramowicz et al. 1988; Madau 1988), 
they cannot illuminate the directions near their 
equatorial plane by the disk self-occultation (Fukue 2000).

As discussed in Paper I, some AGNs 
with presumably super-Eddington accretion rates 
show the weak NIR emission  
(Ark564, TonS180, J0005 and J0303; 
Rodr{\'{\i}}guez-Ardila \& Mazzalay 2006;
Kawaguchi et al. 2004; 
Jiang et al. 2010; see, however, Hao et al. 2010).
Small $\theta_{\rm max}$ 
due to the self-occultation by a geometrically thick disk
can be a reason for the weakness. 
Moreover, the observed data do not support the concept of 
Eddington-limited accretion (Collin \& Kawaguchi 2004).
Thus, a strong anisotropy of the disk emission, 
such as Equation (\ref{eq:flux}) and the disk self-occultation, 
is required to allow gas infall to super-Eddington 
accreting sources.
In this subsection, we examine the influences of the disk 
thickness and the accretion rate upon the NIR emission 
in more detail.

In principle, the disk thickness is a function of the distance 
from the central BH.
Both the illumination spectrum from an AGN disk and the absorption 
efficiency of dust 
have their peaks at Far-UV (e.g., Laor \& Draine 1993).
Therefore, we deduce the disk thickness at the Far-UV emitting region. 
Based on the work 
by Kawaguchi (2003; his Figure 5 for a $10^{6.5} \, M_\odot$ BH),
the semi-thickness of the disk (=\,90\degr$- \, \theta_{\rm max}$) 
at the region with the temperature of (4--5)\,$10^4$\,K 
are 1\degr, 4\degr, 17\degr\ and 39\degr, for the accretion rates of 
1, 10, 100 and 1000\,$L_{\rm Edd}/c^2$, respectively.
Comparing the first and the last cases, 
the solid angle of the torus illuminated by the disk (Equation \ref{eq:omega})
differs by a factor of $\sim$\,9 
(a smaller $\Omega_{\rm torus}$ with a higher accretion rate).

\begin{figure}
\plotone{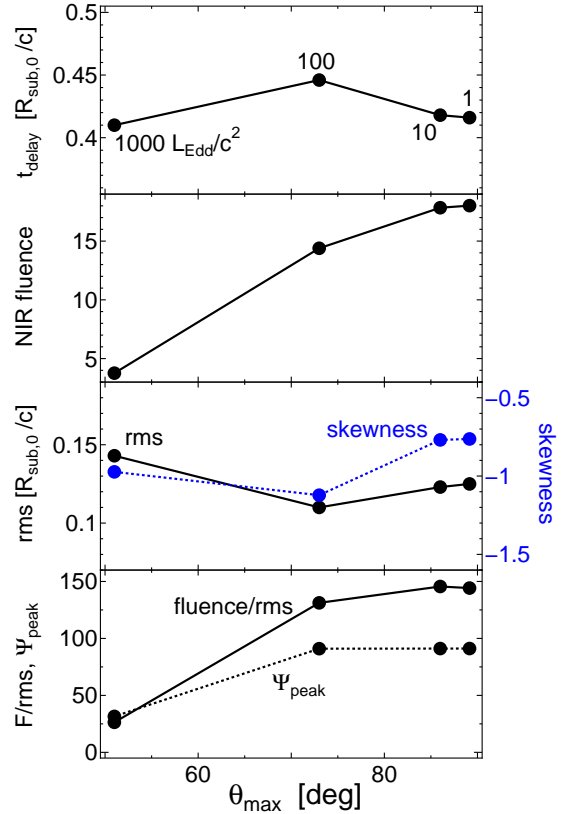}
\caption{
Same as Figures \ref{fig:thetaobs_lssp} and \ref{fig:thetamin_tlss}, 
but for the $\theta_{\rm max}$-dependency.
Corresponding gas accretion rates onto the BH 
in the unit of $L_{\rm Edd}/c^2$ are labelled in the {\it top} panel 
for $t_{\rm delay}$.
{\it Second}:  
Larger accretion rates make the disk thicker 
(i.e., smaller $\theta_{\rm max}$) and the 
shade of the disk itself larger (i.e., smaller $\Omega_{\rm torus}$).
Thus, the NIR fluence becomes small as the accretion rate increases.
{\it Third}: 
$rms$ and $s$ as well as $t_{\rm delay}$ are insensitive to $\theta_{\rm max}$.
{\it Bottom}:  
With high accretion rates, the NIR variability will be 
relatively small.
\label{fig:thetamax_tlss}}
\end{figure}

Transfer functions for four accretion rates are presented 
in Figure \ref{fig:thetamax_tf}.
As we see in Section \ref{sec:theta_contri}, $t_{\rm delay}$ is insensitive 
to $\theta_{\rm max}$ (top panel of Figure \ref{fig:thetamax_tlss}).
Therefore, the observed broad range of $t_{\rm delay}$ is not 
reproduced by a change of the accretion rate. 
Larger accretion rates make the disk thicker and the 
shade of the disk itself larger (i.e., smaller $\Omega_{\rm torus}$; 
Equation (\ref{eq:omega})).
Thus, the NIR fluence becomes small as the accretion rate increases, 
as shown in the second panel. 
By changing the accretion rate from 1 to 1000$L_{\rm Edd}/c^2$, 
the fluence becomes $\sim \! 1/5$.
This is consistent with the weakness of the 
X-ray emission line from neutral iron, which potentially originates 
in the illuminated torus, in a narrow-line quasar 
(Takahashi et al. 2010; cf. Page et al. 2004).
Similar to $t_{\rm delay}$, $rms$ and $s$ are insensitive to 
$\theta_{\rm max}$ (third panel).
On the contrary, 
the height 
of $\Psi (t)$ (bottom panel) 
is affected by 
the accretion rates in the sense that super-Eddington sources 
will show 
a low absolute variation.

Other than the reduction of $\theta_{\rm max}$ and $\Omega_{\rm torus}$, 
super-Eddington accretion rates cause another influence upon 
the NIR flux.
Sub-Eddington accretion disks suffer little from the disk self-gravity,
hence extend far away from the central BH, 
radiating across UV, optical and NIR bands 
(Tomita et al. 2006; Kishimoto et al. 2008).
When the accretion rate becomes super-Eddington, 
the disk self-gravity starts to govern the disk and 
truncate the outer part of the disk.
Due to the truncation, 
super-Eddington disks 
do not radiate at NIR (Kawaguchi et al. 2004).
To sum up, both small $\theta_{\rm max}$ of the torus and 
the small outer radius of the disk, 
caused by the high accretion rate, 
provide less NIR emission.

In contrast to the result for a thin disk with a sub-Eddington
accretion rate of $L_{\rm Edd}/c^2$, 
a super-Eddington accretion rate leads to a much weaker 
NIR emission (with a help of the disk self-gravity) and 
to a low 
time response.

Here, we summarise 
the three dependencies examined above. 
Only the variation of the viewing angles reproduces the observed 
range of the time delay.
Therefore, we propose that the viewing angle is the key parameter 
responsible for the observed scatter 
in the $t_{\rm delay}$--$L_{\rm UV}$ diagram.
A weak NIR emission (such as in hot-dust-poor AGNs in the 
weakest cases) indicates either 
a thin torus, 
a thick torus  
or the super-Eddington accretion.
On the other hand, 
a stronger NIR flux (a redder NIR-optical color) than the fiducial one 
is obtained only by a large viewing angle for a modestly thick torus.

\section{Misaligned Torus} \label{sec:misaligned}

So far, we have assumed that the rotation axis of the torus 
is aligned to that of the disk.
Since a specific angular momentum is likely larger 
in the torus than in the disk,
the assumption sounds plausible.
Indeed, we have examples, where the rotation axes of 
maser disks are aligned 
to the jet directions (e.g., Miyoshi et al. 1995; Meisenheimer et al. 2007; 
Mamyoda et al. 2009).
However, there are also a number of examples showing the misalignment between 
the maser disks and the jets (e.g., Yamauchi et al. 2004; 
Raban et al. 2009).
Random orientation between jets and galactic axes is also reported 
(Clarke et al. 1998; Nagar \& Wilson 1999; Schmitt et al. 2001).

\begin{figure}
\epsscale{.85}
\plotone{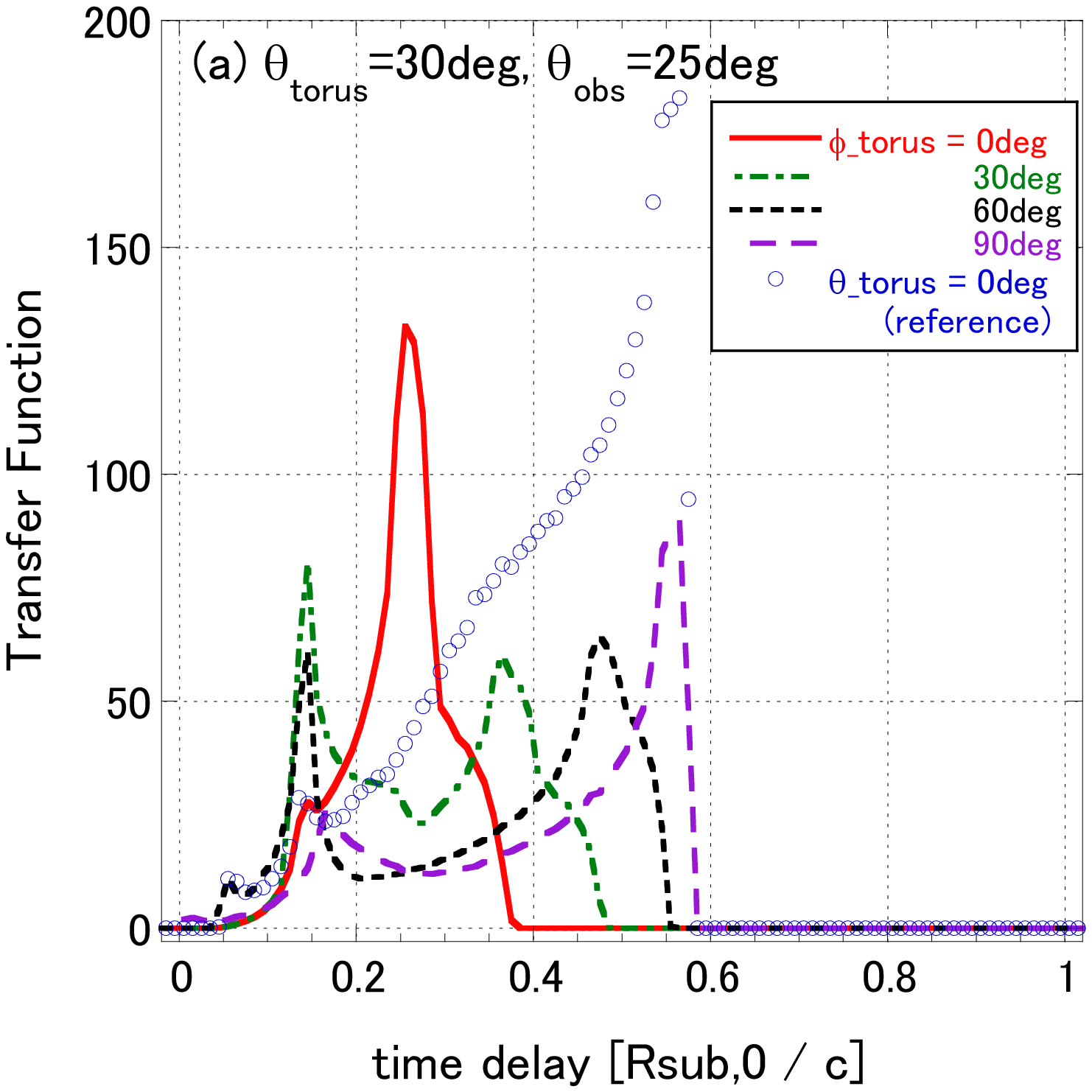}
\plotone{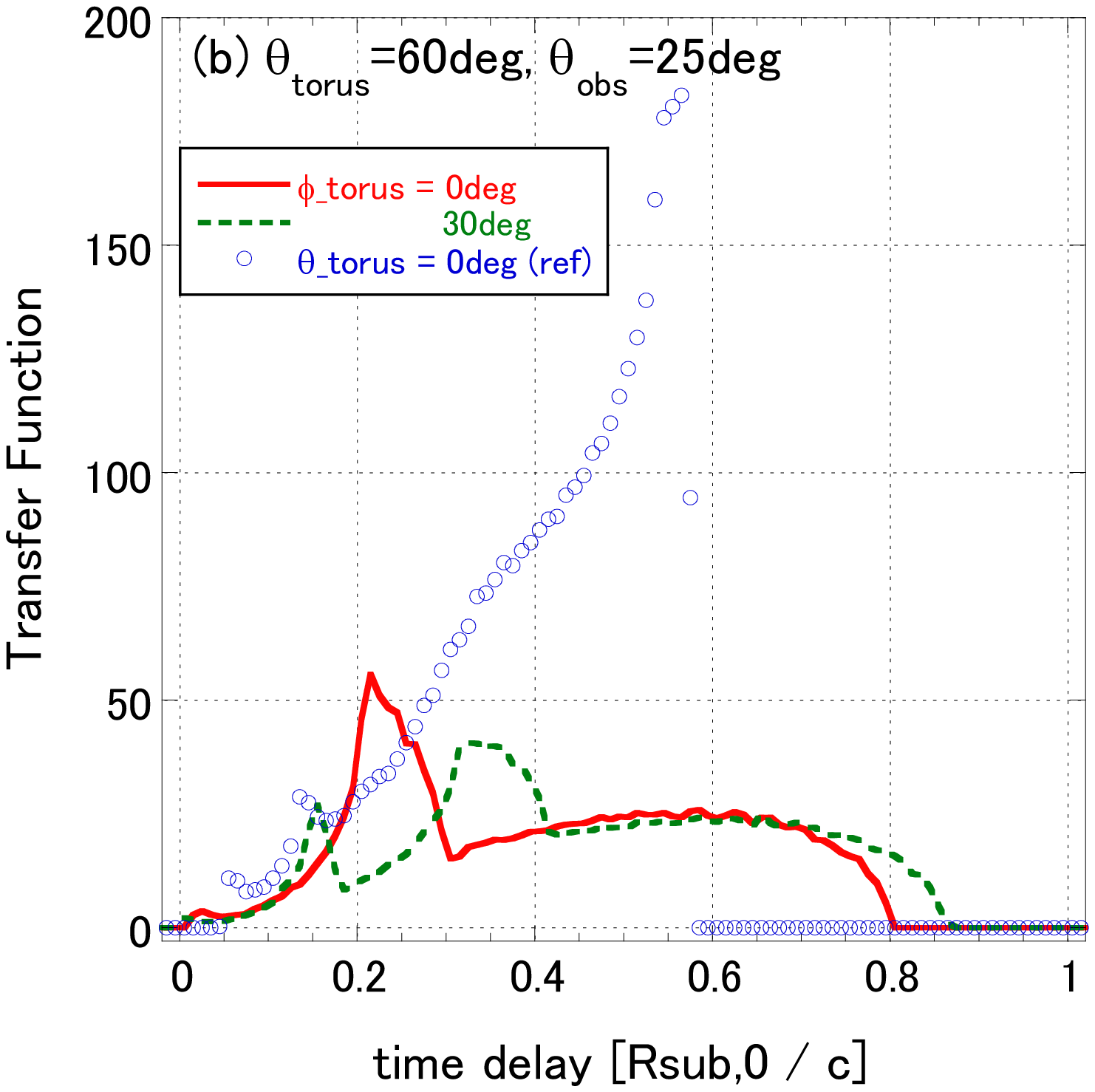}
\plotone{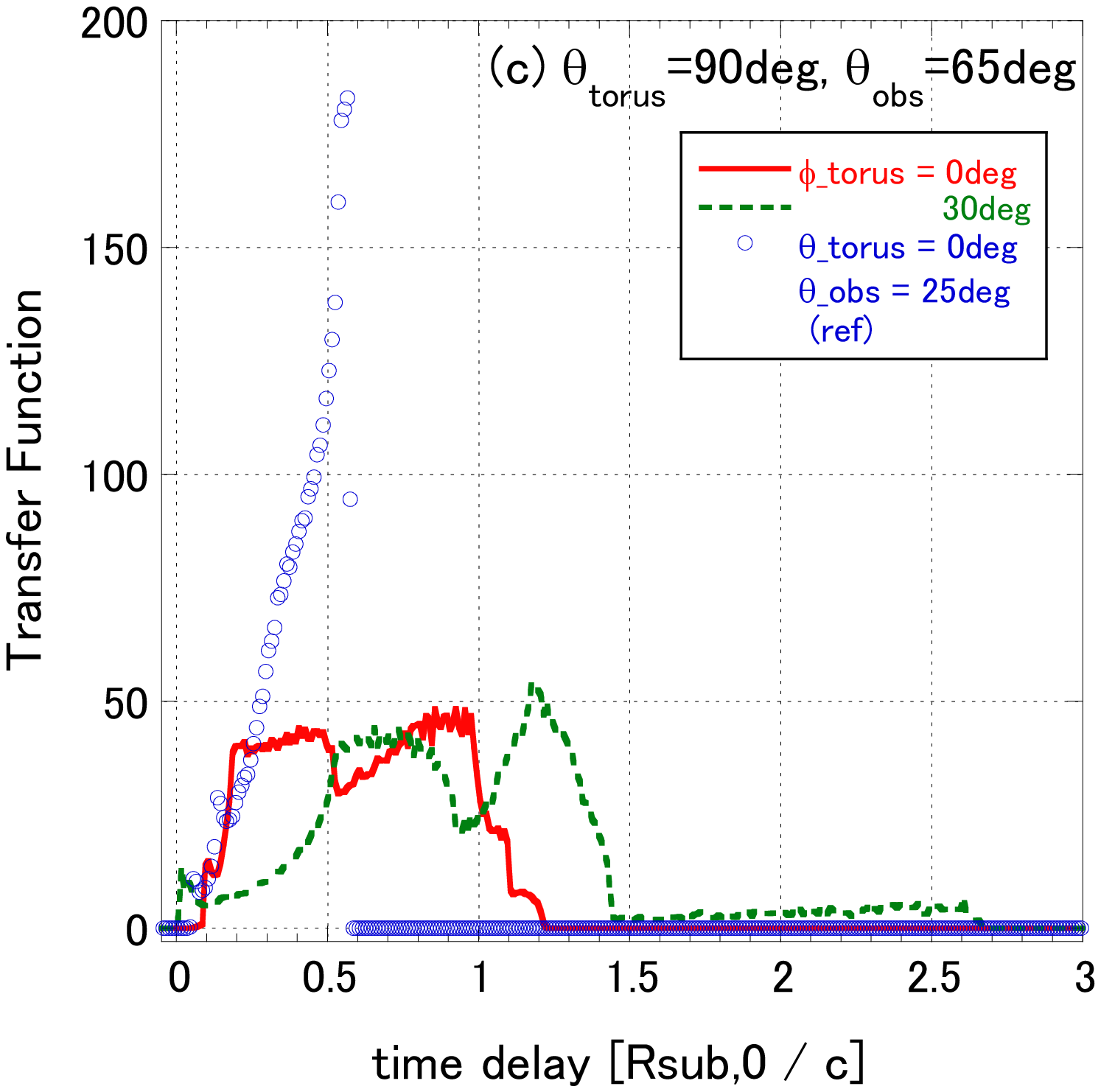}
\caption{Transfer functions for the misaligned tori with 
various $\theta_{\rm torus}$ and $\phi_{\rm torus}$. 
For reference, open blue circles indicate the response of the 
aligned torus seen from $\theta_{\rm obs} = 25$\degr\ (identical 
to the twice enlarged version of 
the blue dotted line in Figure \ref{fig:phi_contribution}). 
(a) $\theta_{\rm torus} = 30$\degr\ and 
(b) $\theta_{\rm torus} = 60$\degr\ with 
$\theta_{\rm obs} = 25$\degr.
(c) $\theta_{\rm torus} = 90$\degr\ with 
$\theta_{\rm obs} = 65$\degr.
In (b) and (c), the NIR emission from $\theta > \pi / 2$, which 
produces the NIR response at a long time delay, is visible.
\label{fig:misalign}}
\end{figure}

In this section, we investigate the consequences of the 
misalignment between the torus and the disk axes.
Now, we introduce $\theta_{\rm torus}$ and $\phi_{\rm torus}$ to 
specify the rotation axis of the torus relative to that of the 
disk and the observer ($\phi = 0$).
Figure \ref{fig:schematic} shows the geometry with 
$\theta_{\rm torus} \approx 20$\degr\ and $\phi_{\rm torus} = 0$\degr.

Three $\theta_{\rm torus}$ are examined; 30\degr, 60\degr\ and 90\degr.
In the first two cases, we adopt $\theta_{\rm obs}$ of $25$\degr.
Since the direction to $\theta_{\rm obs}$ of $25\degr$  
is obscured by the torus if $\theta_{\rm torus} = 90$\degr, 
$\theta_{\rm obs}$ of 65\degr\ is chosen in the last case. 
With $\theta_{\rm torus} = 60$\degr\ and 90\degr, 
the inner edge of the torus at $\theta > \frac{\pi}{2}$ becomes visible.
Thus, for these two $\theta_{\rm torus}$, we integrate $\theta$ not only 
from 0 to $\theta_{\rm max}$ but also from $\pi - \theta_{\rm max}$ to 
$\pi$.
Here, the semi-thickness of the torus and the 
disk are fixed at 45\degr\ and 1\degr\ (i.e., 
$\theta_{\rm min} = 45$\degr\ and $\theta_{\rm max} = 89$\degr), respectively.

Figure \ref{fig:misalign} shows the transfer functions 
(integrating $\phi $ from 0 to $2 \pi$), 
and Table \ref{tbl:misalign} summarises the results of a series of calculations.
For reference, 
we also show the response of the 
aligned torus from $0 \le \phi \le 2\pi$, 
which is identical to the twice enlarged version of 
the dotted line in Figure \ref{fig:phi_contribution} 
(obtained by $0 \le \phi \le \pi$ integration).
For $\theta_{\rm torus}$ of 90$\degr$, the disk illumination flux toward the 
torus is larger than the flux toward the observer, meaning that the 
inner radius of the torus is large.
Blank fields (and $\phi_{\rm torus} \ge 120$\degr) 
in the Table mean that the direction to the observer 
from the central BH is obscured by the torus with such parameter sets.
In other words, a large $\phi_{\rm torus}$ 
excludes type-1 AGNs, 
and is biased to type-2 AGNs.

\begin{table}
\begin{center}
\caption{Results for Misaligned Tori.\label{tbl:misalign}}
\begin{tabular}{ccccc}
\tableline
\tableline
\multicolumn{5}{c}{$\theta_{\rm obs} = 25$\degr}\\
\tableline
$\theta_{\rm torus}$ [deg] & \multicolumn{4}{c}{$\phi_{\rm torus}$ [deg]}\\
  & 0 & 30 & 60 & 90\\
\tableline
0 & \multicolumn{4}{c}{$t_{\rm delay} = 0.42$ ($R_{\rm sub,0}/c$)\tablenotemark{a}}\\
  & \multicolumn{4}{c}{Fluence = 36\tablenotemark{b}}\\
  & \multicolumn{4}{c}{$rms$ = 0.13}\\
  & \multicolumn{4}{c}{$s$ = $-0.76$}\\
\tableline
30 & 0.25  &  0.28 &  0.35 & 0.41  \\
   & 13    &  13   &  13   & 12    \\
   & 0.06  &  0.11 &  0.15 & 0.15  \\
   &$-0.40$&$-0.04$&$-0.53$&$-0.75$\\
\tableline
60 & 0.43 &  0.47 &  \nodata & \nodata\\
   & 16   &  17   &   & \\
   & 0.19 &  0.20 &   & \\
   &$0.11$& $0.03$&   & \\
\tableline
\multicolumn{5}{c}{$\theta_{\rm obs} = 65$\degr}\\
\tableline
90 & 0.62  &  0.98 &  \nodata & \nodata\\
   & 38    &  44   &   & \\
   & 0.28  &  0.50 &   & \\
   &$0.003$& $1.0$ &   & \\
\tableline
\end{tabular}
\tablenotetext{a}{Hereafter, the time delay is quoted in this unit.}
\tablenotetext{b}{NIR fluence for the fiducial parameter set is 
calculated by integrating $\theta$ from 0 to $2 \pi$, which 
has twice the fluence shown in 
Sections \ref{sec:model} and \ref{sec:aligned}.}

\tablecomments{
Blank fields (and $\phi_{\rm torus} \ge 120$\degr) 
mean that the direction to the observer (at $\phi =0$)
from the central BH is obscured by the torus with such parameter sets.
}

\end{center}
\end{table}

A variety of $t_{\rm delay}$ is achieved from 
0.25 to 0.98\,$R_{\rm sub,0}/c$, which 
is wide enough to match up with the observed range 
(Figure \ref{fig:thetaobs_tdelay}).
A short $t_{\rm delay}$ is 
obtained for $\theta_{\rm torus}$ of 30$\degr$, 
and is associated with a small 
NIR fluence.
The geometry with $\theta_{\rm torus} = 30$\degr\ and 
$\phi_{\rm torus} =0$\degr\ is similar to a pole-on 
view 
of aligned tori (Section \ref{sec:qobs}), 
producing a very narrow NIR response at a short time delay. 
On the other hand, 
a large $t_{\rm delay}$ is realized for 
$\theta_{\rm torus}$ of 90$\degr$, 
accompanied 
with a fluence similar to the one for the fiducial parameter.

As the torus becomes inclined relative to the disk, 
the torus hides not only 
the region with a short delay but also various area 
of the inner edge of the torus, 
making the NIR fluence small and the time response complicated.
Contributions from $\theta > \frac{\pi}{2}$, which are observable for 
 $\theta_{\rm torus}$ of $60\degr$ and $90\degr$, 
 appear at a longer time delay.
Thus, they tend to increase the time delay, the NIR fluence and the 
skewness ($s \ga 0$).
With $\theta_{\rm torus} = 90$\degr, 
the NIR emission from $\theta > \frac{\pi}{2}$ is large enough to 
compensate the flux reduction due to the torus self-occultation, 
exhibiting a comparable fluence to the 
reference result.
A long $t_{\rm delay}$ ($\sim R_{\rm sub,0}/c$), 
if observed in the future, may be a signature of a large 
$\theta_{\rm torus}$ ($\sim \! 90$\degr; i.e., heavily misaligned).

In case the torus self-occultation does not work due to an extremely 
small volume filling factor of clumps etc (Appendix), 
an aligned torus shows the NIR response as described in 
Section \ref{sec:theta_contri} with $t_{\rm delay} = 0.78 \,R_{\rm sub,0}/c$.
Summing up all the four lines in Figure \ref{fig:theta_contribution}b 
and multiply the fluence by two (to convert the $0 \le \phi \le \pi$ 
integration to the 0--2$\pi$ integration), 
we obtain the fluence of 140.
In contrast, we find that misaligned tori without the self-occultation 
effect show a yet longer delay with a huge NIR fluence.
For instance, 
for $\theta_{\rm torus} = 60\degr$ with 
$\theta_{\rm obs} = 25\degr$ and 
for $\theta_{\rm torus} = 90\degr$ with 
$\theta_{\rm obs} = 65\degr$, 
$t_{\rm delay} = 1.2$ and $1.9\, R_{\rm sub,0}/c$ with the 
fluence of 250 and 960, respectively.
Therefore, an extremely long ($> R_{\rm sub,0}/c$) delay with 
a big NIR flux would mean a misaligned optically thin torus.

\section{Summary} \label{sec:summary}

According to recent models, 
the accretion disk and the BH in AGNs are surrounded by 
a clumpy torus, 
with its inner radius governed by the dust sublimation
process.
Regarding the inner radius of the torus,
there was a systematic deviation between the observational 
results and the theory.
In Paper I, we showed that the anisotropy of the disk emission 
resolves this 
conflict for 
a typical type-1 AGN. 
We found that the anisotropy makes the torus inner region 
closer to the central BH and concave.
Furthermore, the innermost edge of the torus may connect 
with the outermost edge of the accretion disk continuously.

In this study, 
we have calculated the NIR flux variation of the torus in response 
to a UV flash for 
various geometries of the disk, the torus and the observer. 
Anisotropic illumination by the disk and the effect of 
the torus self-occultation contrast our study with earlier works.
We have found that 
both the waning effect of each clump and the torus self-occultation 
selectively reduce the emission from  the region with a short delay.
Thus, the resultant NIR time response shows a 
$\theta_{\rm obs}$-dependent delay and 
an asymmetric profile with a negative skewness, 
opposing to the results for optically thin tori (Barvainis 1992). 

By contrast with the fiducial viewing angle 
of 25\degr, 
a small viewing angle results in a short time delay 
with a narrow and peaky response.
On the other hand, a more inclined viewing angle leads to 
a longer delay with a broader profile and 
to 
a redder NIR-to-optical color.
We propose that the red NIR-optical color 
of type-1.8/1.9 objects is caused by not only 
the dust extinction but also intrinsically red color. 
The computed range of $t_{\rm delay}$ coincides with 
the observed one.

Compared with the modest torus thickness of $45\degr$, 
both a thick and a thin tori display the weaker NIR emission, 
consistent with the work by Nenkova et al. (2008).
In other words, a modest thickness of the torus 
leads to the strongest NIR emission. 
A selection bias is thus expected such that NIR-selected 
AGNs tend to possess moderately thick tori.
For thin tori, 
we have found that the NIR fluence is in proportion 
to $\Omega_{\rm torus}^{1.9}$.
As the torus becomes thicker,
the NIR response shows a slightly longer delay 
with a narrower and more heavily skewed profile 
due to the torus self-occultation. 
This trend opposes to the 
viewing angle dependency, where 
the delay and the width are positively correlated each other.
On the contrary, 
 as the torus gets thinner, 
the NIR response becomes more rapid, narrower,  
closer to time-symmetric and low. 

In contrast to 
a thin disk with a sub-Eddington
accretion rate, 
a super-Eddington accretion rate leads to a much weaker ($< \! 1/5$)
NIR emission due to the disk self-occultation and 
the disk truncation by the self-gravity, and 
to a low 
time response.

Among the three dependencies examined for aligned tori, 
only the variation of the viewing angles reproduces the observed range
of the delay.
Therefore, we propose that the viewing angle is the key parameter 
responsible for the observed scatter 
about the regression line in the $t_{\rm delay}$ v.s. $L_{\rm UV}$ diagram.
Conversely, the measurements of $t_{\rm delay}$ 
are potentially useful 
to estimate the inclination angles.

We have also investigated the consequences of the 
misalignment between the torus and the disk axes.
A variety of $t_{\rm delay}$ is achieved, which 
is wide enough to cover the observed range.
A short delay 
is obtained for a small misalignment and is 
associated with a small NIR fluence, 
while a 
a long delay is obtained 
for a largely misaligned torus 
with an usual fluence. 
This trend contrasts with the 
viewing angle dependency for aligned tori, 
where a short delay is associated 
with a normal NIR fluence while a long delay 
means a large NIR fluence.
For 
highly misaligned cases, 
contributions from $\theta > \frac{\pi}{2}$
 increase the time delay (up to $\sim R_{\rm sub,0}/c$), the NIR fluence and the 
skewness ($s \ga 0$).

In case the torus is optically thin (with an inefficient 
self-occultation), 
the time delay of the NIR emission from an aligned torus 
becomes longer.
Moreover, 
misaligned optically thin tori 
show a yet longer delay ($> R_{\rm sub,0}/c$) with a huge NIR flux.

From an observational point of view, 
these numerical results are summarised as follows.
If the observed time delay of the NIR emission is 
short, it will mean either a small viewing angle, 
a geometrically thin torus, or a slightly misaligned torus.
If the delay is long, on the other hand, 
it indicates either an inclined viewing angle, 
an aligned optically thin torus, or a largely misaligned torus.
An extremely long delay ($> R_{\rm sub,0}/c$) 
would mean a misaligned optically thin torus.
As to the NIR flux, 
a blue NIR-to-optical color 
(i.e., a weak NIR emission such as in 
hot-dust-poor AGNs in the weakest cases)
indicates 
either 
a geometrically thin torus, 
a geometrically thick torus, 
a super-Eddington accretion, 
or 
a slight misalignment between the torus and the disk.
On the other hand, 
a red NIR-optical color (a large NIR emission) means 
 a large viewing angle with a modest geometrical thickness of the torus 
 or a largely misaligned optically thin torus.

\acknowledgments

We are grateful to 
T.~Minezaki, 
M.~Umemura, 
M.~Shirahata, 
T.~Nakagawa, 
M.~Gaskell, 
J.~Fukue, 
H.~Takahashi, 
Y.~Miki, 
M.~Kishimoto 
and
S.~Koshida  
for useful discussions, 
and the anonymous referee for helpful comments.
This work was partly supported by the Grants-in-Aid of the Ministry
of Education, Science, Culture, and Sport (19740105, 21244013).

\appendix
\section{Optical Thickness of Clumpy Tori} \label{sec:clumpy_tori}

We briefly summarise the basic properties of clumps in the torus.
The main aim here is to deduce the optical thickness of the clumpy torus.
If it is optically thin at NIR (as Barvainis 1992 assumed), 
we need to consider the NIR emission from all the clumps 
including those at the 
far side of the torus which is illuminated by the 
back side (relative to the observer) of the accretion disk.
In contrast, if it is optically thick, 
the torus self-occultation should be incorporated appropriately.

Each clump 
located at a distance of $r$ from 
the central BH is characterised by the radius of $R_{\rm clump}$ and 
the mass of $M_{\rm clump}$. 
Clumps that survive against its own thermal pressure and tidal 
shearing by the BH must be heavier than the Jeans mass and/or 
more compact than the tidal (Hill or Roche) radius
(e.g., Vollmer et al. 2004). 
Marginally stable clumps, at the boundary of these criteria, 
have the following various quantities 
(H\"{o}nig \& Beckert 2007), 
\begin{eqnarray}
R_{\rm clump} &=& \frac{\sqrt{\pi} c_s r^{1.5}}{3\sqrt{ G M_{\rm BH}}}
 \approx 0.01 \,\mathrm{pc}
 \brfrac{c_s}{3 \, \mathrm{km}\,\mathrm{s}^{-1}}
 \brfrac{r}{\mathrm pc}^{1.5}
 \brfrac{M_{\rm BH}}{10^7 \,M_\odot}^{-0.5}, \label{eq:clumpsize} \\
M_{\rm clump} &=& \frac{\pi c_s^2}{3 G} R_{\rm clump}
 \approx 20 \,M_\odot 
 \brfrac{c_s}{3 \, \mathrm{km}\,\mathrm{s}^{-1}}^3
 \brfrac{r}{\mathrm pc}^{1.5}
 \brfrac{M_{\rm BH}}{10^7 \,M_\odot}^{-0.5}, \\\label{eq:clumpmass}
\rho_{\rm clump} 
 &=& \frac{M_{\rm clump}}{\frac{4}{3} \pi R_{\rm clump}^3}
 \approx  5 \, 10^{-16} \,\mathrm{g}\,\mathrm{cm}^{-3}
 \brfrac{r}{\mathrm pc}^{-3}
 \brfrac{M_{\rm BH}}{10^7 \,M_\odot},\label{eq:clumpdensity}\\
N_{\rm H, clump} 
 &\approx& \frac{\rho_{\rm clump}}{m_\mathrm{p}}\, R_{\rm clump}
 \approx 8\,10^{24}\,\mathrm{cm}^{-2} 
 \brfrac{c_s}{3 \, \mathrm{km}\,\mathrm{s}^{-1}}
 \brfrac{r}{\mathrm pc}^{-1.5}
 \brfrac{M_{\rm BH}}{10^7 \,M_\odot}^{0.5}.\label{eq:clumnh}
\end{eqnarray}
Here, $\rho_{\rm clump}$, $N_{\rm H, clump}$, $c_s$ and $m_\mathrm{p}$ 
are the mean 
density in the clump, the column density 
of the clump,  the sound speed in the clump and the proton mass, respectively.
The normalization in $\rho_{\rm clump}$  
(corresponding to $3 \, 10^8$\,cm$^{-3}$) 
is consistent with the observed lower limit 
for the mean number density 
($\ga \! 10^7$\,cm$^{-3}$; Geballe et al. 2006; Shirahata et al. 2007).
Incidentally, clumps in the broad line region 
(at $r \sim 10^{16}$\,cm) have their 
sizes around 10$^{13}$\,cm (Risaliti et al. 2009; Maiolino et al. 2010), 
consistent with the extrapolation of Equation (\ref{eq:clumpsize}) to smaller $r$.
However, their density ($\sim 10^{11} \mathrm{cm}^{-3}$) and $N_{\rm H, clump} $
[$\sim (2-9)\, 10^{23} \mathrm{cm}^{-2}$] are much less than those expected from 
Equations (\ref{eq:clumpdensity}) and (\ref{eq:clumnh}).

The column density above means Compton-thick 
($\sim \! 5 \, \sigma_\mathrm{Thomson}^{-1}$),
and corresponds to optical depths at $V$- and $K$-bands of 
1400 and 160, 
respectively, using the conventional extinction law 
(Savage \& Mathis 1979; Cardelli et al. 1989). 
In radiative transfer calculations of AGN tori, 
clumps with optical depth at $V$-band of $30$--$100$ are 
often adopted  
(Nenkova et al. 2008; H\"{o}nig et al. 2008; Deo et al. 2011).
Krolik \& Begelman (1988) estimates $N_{\rm H, clump}$ 
to be $\sim \! 7 \, 10^{23}\,\mathrm{cm}^{-2} $ 
($\sim$\,one tenth of Equation (\ref{eq:clumnh})).
Even with such a reduction of the clump opacity by 
$\frac{1}{10}$--$\frac{1}{40}$, 
each clump is still opaque 
to NIR photons.

Therefore, optical thickness of the torus $\tau_{\rm torus}$ is 
simply related to the 
probability that the incoming rays hit clumps.
We treat the two directions separately:
one in the vertical direction (parallel to the rotation axis) 
with a suffix of $\parallel $, 
and another along the equatorial plane with a suffix of $\perp $.
 The key parameter here is the volume filling factor 
of clumps in the torus $f$, for which we assume 0.03 (Vollmer et al. 2004).
By writing the number density of clumps in the torus by $n_{\rm c}$, 
\begin{eqnarray}
n_{\rm c} &=& \frac{f}{\frac{4}{3} \pi R_{\rm clump}^3},\\
\tau_{\rm torus, \parallel } &=& n_{\rm c} \, \pi R_{\rm clump}^2 \, H, \\
 &\approx& 0.02 \brfrac{r}{R_{\rm clump}} \brfrac{f}{0.03},
\end{eqnarray}
where we assume the thickness of the torus $H$ 
to be $\sim \! r$. 
Now, $\tau_{\rm torus}$ means the mean number of clumps along the ray, 
and 
a fraction $e^{-\tau_{\rm torus}}$ of the incoming rays 
 pass through the torus without encountering any clumps 
(Natta \& Panagia 1984; Nenkova et al. 2002).

\begin{figure}
\epsscale{0.45}
\plotone{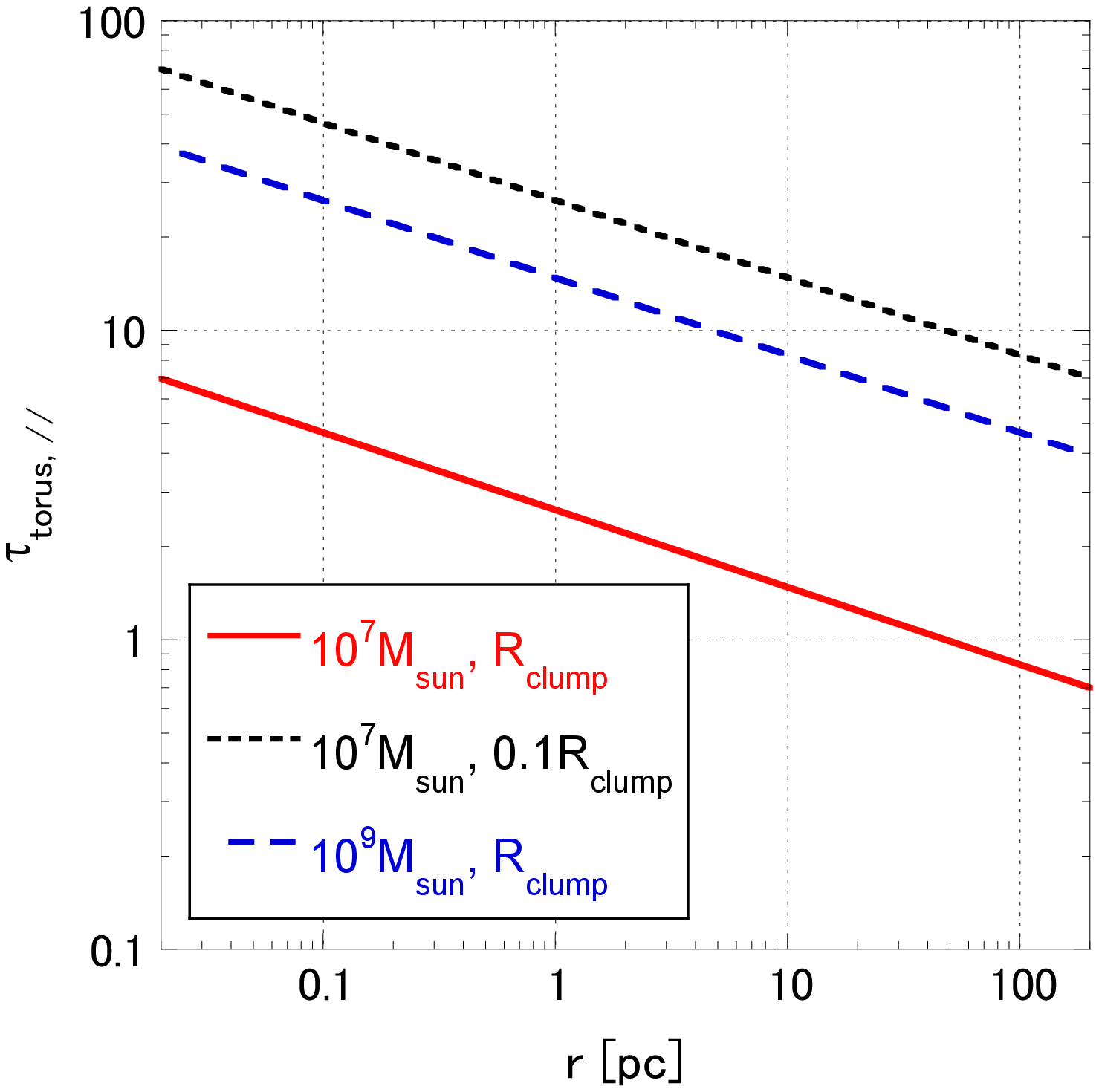}
\caption{
Optical thickness of the clumpy torus 
in the vertical direction 
as a function of 
the distance from the central BH $r$.
Red solid and black dotted lines are for $M_{\rm BH}$ of $10^7 M_\odot$, 
with the clump radius assumed to be one tenth of that 
in Equation (\ref{eq:clumpsize}) in the latter.
Blue dashed line represents the case of 
$10^9 M_\odot$.
For all the cases, 
the inner part of the clumpy torus is optically thick.
\label{fig:tautorus}}
\end{figure}

If we adopt the clump size in Equation (\ref{eq:clumpsize}), then 
\begin{equation}
\tau_{\rm torus, \parallel } = 
3 
 \brfrac{c_s}{3 \, \mathrm{km}\,\mathrm{s}^{-1}}^{-1}
 \brfrac{r}{\mathrm pc}^{-0.5}
 \brfrac{M_{\rm BH}}{10^7 \,M_\odot}^{0.5} \brfrac{f}{0.03}.
\end{equation}
Assuming that clumps are mainly heated by the direct illumination from 
the central accretion disk (i.e. ignoring the irradiation from 
nearby clumps), the clump temperature will be proportional
to $r^{-0.5} L^{0.25}$, with $L$ being the illumination luminosity 
(cf. Nenkova et al. 2008).
Therefore, the size of torus emission is larger 
for larger $L$ at longer $\lambda$, 
in the form $\propto \lambda^2 L^{0.5}$ (Tristram et al. 2009).
Then, we assume for $c_s$ as follows,
\begin{equation}
c_s = c_0 
 \brfrac{r}{\mathrm pc}^{-\frac{1}{4}}
 \brfrac{L}{L_{\rm Edd}}^{\frac{1}{8}}
  \brfrac{M_{\rm BH}}{10^7 \,M_\odot}^{\frac{1}{8}},
\end{equation}
with $L_{\rm Edd}$ being the Eddington luminosity.
In summary, 
\begin{equation}
\tau_{\rm torus, \parallel } = 
3
 \brfrac{c_0}{3 \, \mathrm{km}\,\mathrm{s}^{-1}}^{-1}
 \brfrac{r}{\mathrm pc}^{-\frac{1}{4}}
 \brfrac{L}{L_{\rm Edd}}^{-\frac{1}{8}}
 \brfrac{M_{\rm BH}}{10^7 \,M_\odot}^{\frac{3}{8}} 
 \brfrac{f}{0.03}.\label{eq:tautorus}
\end{equation}
Figure \ref{fig:tautorus} shows the 
optical thickness of the clumpy torus as a function of 
$r$.
Throughout, we fix $c_0$, $L / L_{\rm Edd}$ and $f$ so that 
these in parentheses in Equation (\ref{eq:tautorus}) equal unity.
For instance, $\tau_{\rm torus, \parallel }$ is 
about five
at $r$ of 0.1pc.

The radius where the torus is opaque $r_{\rm opaque}$ 
is achieved by setting $\tau_{\rm torus, \parallel }=1$:
\begin{equation}
r_{\rm opaque}
= 50 \,\mathrm{pc} 
 \brfrac{c_0}{3 \, \mathrm{km}\,\mathrm{s}^{-1}}^{-4}
 \brfrac{L}{L_{\rm Edd}}^{-0.5}
 \brfrac{M_{\rm BH}}{10^7 \,M_\odot}^{1.5}
 \brfrac{f}{0.03}^{4}
\end{equation}
The inner part within this radius of the 
torus is 
likely optically thick.
In this study, we therefore consider basically emission 
from the near side of the torus, as we did in Paper I.
This assumption can be tested in principle via 
the profile of broad emission lines and its time variation 
(e.g., Peterson 2001).
If some indications of the emission from the far side 
of the torus is observed, 
the volume filling factor seems extremely low 
($\ll \! 0.03/5$) 
or the torus is very thin 
($H/r \ll 1/5$).

Similarly, we deduce the optical thickness perpendicular to the 
rotation axis $\tau_{\rm torus, \perp }$, 
which is equivalent to $\mathcal{N}_\mathrm{0}$ 
(the average number of clumps along radial equatorial rays) denoted 
by Nenkova et al. (2008).
They showed that 
$\mathcal{N}_\mathrm{0}$ is likely between 5 and 15, 
consistent with our estimation below.
Adopting the inner radius of $R_{\rm sub,0}$ and 
the clump size in 
Equation (\ref{eq:clumpsize}), 
\begin{eqnarray}
\tau_{\rm torus, \perp } &=& 
\int_{R_{\rm sub,0}} n_{\rm c} \, \pi R_{\rm clump}^2 \, dr, \\
 &\approx& 
8
 \brfrac{c_0}{3 \, \mathrm{km}\,\mathrm{s}^{-1}}^{-1}
 \brfrac{M_{\rm BH}}{10^7 \,M_\odot}^{0.5} 
 \brfrac{f}{0.03}
 \brfrac{T_{\rm sub}}{1500 \,\mathrm{K}}^{-0.7}
 \brfrac{a}{0.05 \,\mu \mathrm{m}}^{-\frac{1}{8}}.
\end{eqnarray}
Here, we assume $L = 2.5 \, L_{\rm UV}$.
Since this optical thickness is also larger than unity, 
we restrict ourselves to non-obscured objects
(i.e., the geometry where the line of 
sight to the central BH is not blocked by the torus).

\end{document}